\documentstyle{mn}
\input{psfig}

\newif\ifAMStwofonts

\newcommand{\tF}{\tilde{F}}
\newcommand{\tV}{\hat{V}}
\newcommand{\tA}{\tilde{A}}

\newcommand{\avg}[1]{\left\langle{#1}\right\rangle}

\renewcommand{\r}{\mbox{\boldmath  $r$}}
\renewcommand{\u}{\mbox{\boldmath $u$}}

\newcommand{\ie}{{${\rm i.e.\ }$}}

\renewcommand{\bar}{\overline }

\newcommand{\xiav}{\bar{\xi}}

\ifoldfss
  \ifCUPmtlplainloaded \else
    \NewTextAlphabet{textbfit} {cmbxti10} {}
    \NewTextAlphabet{textbfss} {cmssbx10} {}
    \NewMathAlphabet{mathbfit} {cmbxti10} {} 
    \NewMathAlphabet{mathbfss} {cmssbx10} {} 
  \fi
  \ifAMStwofonts
    \ifCUPmtlplainloaded \else
      \NewSymbolFont{upmath} {eurm10}
      \NewSymbolFont{AMSa} {msam10}
      \NewMathSymbol{\upi}     {0}{upmath}{19}
      \NewMathSymbol{\umu}     {0}{upmath}{16}
      \NewMathSymbol{\upartial}{0}{upmath}{40}
      \NewMathSymbol{\leqslant}{3}{AMSa}{36}
      \NewMathSymbol{\geqslant}{3}{AMSa}{3E}

      \let\leq=\leqslant 
      \let\geq=\geqslant \let\ge=\geqslant
    \fi
  \fi
\fi 

\ifnfssone
  \newmathalphabet{\mathit}
  \addtoversion{normal}{\mathit}{cmr}{m}{it}
  \addtoversion{bold}{\mathit}{cmr}{bx}{it}
  \newmathalphabet{\mathbfit} 
  \addtoversion{normal}{\mathbfit}{cmr}{bx}{it}
  \addtoversion{bold}{\mathbfit}{cmr}{bx}{it}
  \newmathalphabet{\mathbfss} 
  \addtoversion{normal}{\mathbfss}{cmss}{bx}{n}
  \addtoversion{bold}{\mathbfss}{cmss}{bx}{n}
  \ifAMStwofonts
    \ifCUPmtlplainloaded \else
      \UseAMStwoboldmath
      \makeatletter
      \new@mathgroup\upmath@group
      \define@mathgroup\mv@normal\upmath@group{eur}{m}{n}
      \define@mathgroup\mv@bold\upmath@group{eur}{b}{n}
      \edef\UPM{\hexnumber\upmath@group}
      \new@mathgroup\amsa@group
      \define@mathgroup\mv@normal\amsa@group{msa}{m}{n}
      \define@mathgroup\mv@bold\amsa@group{msa}{m}{n}
      \edef\AMSa{\hexnumber\amsa@group}
      \makeatother
      \mathchardef\upi="0\UPM19
      \mathchardef\umu="0\UPM16
      \mathchardef\upartial="0\UPM40
      \mathchardef\leqslant="3\AMSa36
      \mathchardef\geqslant="3\AMSa3E

      \let\leq=\leqslant 
      \let\geq=\geqslant \let\ge=\geqslant
    \fi
  \fi
\fi 

\ifnfsstwo
  \DeclareMathAlphabet{\mathbfit}{OT1}{cmr}{bx}{it}
  \SetMathAlphabet\mathbfit{bold}{OT1}{cmr}{bx}{it}
  \DeclareMathAlphabet{\mathbfss}{OT1}{cmss}{bx}{n}
  \SetMathAlphabet\mathbfss{bold}{OT1}{cmss}{bx}{n}
  \ifAMStwofonts
    \ifCUPmtlplainloaded \else
      \DeclareSymbolFont{UPM}{U}{eur}{m}{n}
      \SetSymbolFont{UPM}{bold}{U}{eur}{b}{n}
      \DeclareSymbolFont{AMSa}{U}{msa}{m}{n}
      \DeclareMathSymbol{\upi}{0}{UPM}{"19}
      \DeclareMathSymbol{\umu}{0}{UPM}{"16}
      \DeclareMathSymbol{\upartial}{0}{UPM}{"40}
      \DeclareMathSymbol{\leqslant}{3}{AMSa}{"36}
      \DeclareMathSymbol{\geqslant}{3}{AMSa}{"3E}

      \let\leq=\leqslant 
      \let\geq=\geqslant \let\ge=\geqslant
    \fi
  \fi
\fi 

\ifCUPmtlplainloaded \else
  \ifAMStwofonts \else 
    \def\upi{\pi}
    \def\umu{\mu}
    \def\upartial{\partial}
  \fi
\fi

\title{Effects of Sampling on Statistics of Large Scale Structure}
\author[S. Colombi et al.]{St\'ephane Colombi,$^{1,2}$ Istv\'an Szapudi$^3$ and Alexander
S. Szalay$^4$\\
$^1$CITA, 60 St George St., Toronto, ON, Canada, M5S 1A7\\
$^2$Institut d'Astrophysique de Paris, CNRS, 98bis bd Arago,
F-75014 Paris, France\\
$^3$NASA/Fermilab Astrophysics Center, Fermi National
Accelerator Laboratory, Batavia, IL 60510-0500\\
$^4$Department of Physics and Astronomy, The Johns Hopkins
University, Baltimore, MD 21218} 
\date{Accepted for publication in MNRAS}
\pagerange{\pageref{firstpage}--\pageref{lastpage}}
\pubyear{1997}
\begin{document}
\maketitle
\label{firstpage}
\begin{abstract}
The effects of sampling are investigated on measurements of 
counts-in-cells in three-dimensional magnitude limited
galaxy surveys, with emphasis on moments of the underlying smooth 
galaxy density field convolved with a spherical window.
A new estimator is proposed for measuring the $k$-th
order moment $\langle \rho^k \rangle$: the weighted factorial moment 
$\tF_k[\omega]$.  Since these statistics are corrected for 
the effects of the varying selection function, they can extract the
moments in one pass without the need of constructing a series
of volume limited samples.

The cosmic error on the measurement of $\tF_k[\omega]$ is computed via the
the formalism of Szapudi \& Colombi (1996), which is
generalized to include the effects of the selection function.
The integral equation for finding the minimum variance weight 
is solved numerically, 
and an accurate and intuitive analytical approximation is derived
$\omega_{\rm optimal}(\r) \propto 1/\Delta(r)$, where 
$\Delta(r)$ is the cosmic error as a function of the distance from the
observer. The resulting estimator is more accurate than the traditional
method of counts-in-cells in volume limited samples, which discards
useful information. As a practical example, it is demonstrated
that, unless unforeseen systematics will prevent it, the proposed method
will extract moments of the galaxy distribution in the future Sloan 
Digital Sky Survey (hereafter SDSS) with accuracy of order few percent
for $k=2$, $3$ and better than $10\%$ for $k=4$ in the scale range of
$1\ h^{-1} \ {\rm Mpc} \leq \ell \leq 50\ h^{-1}$ Mpc. In the particular
case of the SDSS, a homogeneous (spatial) weight $\omega=1$ is
reasonably close to optimal.

Optimal sampling strategies for designing magnitude limited 
redshift surveys are investigated as well. The arguments of
Kaiser (1986) are extended to higher order moments, 
and it is found that the optimal strategy depends greatly on the
statistics and scales considered. A sampling rate $f \sim 1/3-1/10$ is
appropriate to measure low-order moments with $k \leq 4$ in the
scale range $1\ h^{-1} \la \ell \la 50 \ h^{-1}$ Mpc. 
However, the optimal sampling rate increases with $k$, the order considered,
and with $1/\ell$. Therefore count-in-cells statistics in general, 
such as the shape of the distribution function, high order moments, 
cluster selection, etc., require full sampling, 
especially at small, highly nonlinear scales $\ell \sim 1\ h^{-1}$ Mpc. 

Another design issue is the optimal geometry of a catalog, when
it covers only a small fraction of the sky. 
Similarly as Kaiser (1996), we find that a survey composed of 
several compact subsamples of angular size $\Omega_{\rm F}$ 
spread over the sky on a glass-like structure would do better, with
regards to the cosmic error, than the compact or the traditional slice 
like configurations, at least at small scales.
The required dynamic range of the measurements determines
the characteristic size of the subsamples. It is however
difficult to estimate, since an accurate and cumbersome calculation
of edge effects would be required at scales comparable to the size 
of a subsample.
\end{abstract}
\begin{keywords}
large scale structure of the universe --
galaxies: clustering -- methods: numerical -- methods: statistical
\end{keywords}
%
%
\section{Introduction}
%
%
The large scale structure of the Universe is generally admitted to be
homogeneous at scales above $\sim 150$ Mpc. At smaller scales, observations
of the galaxy distribution show a remarkable clustering as evidenced
by  voids, clusters, filaments, and superclusters. According
to standard theories, these structures grew from small 
initial fluctuations under the influence of gravity.
This, together with possible biasing, resulted in the random point 
process represented by galaxies. Thus, statistical methods can be
applied efficiently to galaxy surveys to constrain models of large
scale structure formation. Once the statistical tool is selected, two 
important questions influence its applicability. First, an optimal
{\em sampling strategy} can be used to build a galaxy catalog
(e.g.~Kaiser 1986, hereafter K86), thus maximizing the information
content with respect to the statistics used. Second, an optimal
measurement method can be used to extract the maximum amount of
information present in the catalog. The aim of this work is to address
both of these questions in a quantitative way, focusing on low order 
moments of the probability distribution function (PDF) of the large 
scale galaxy density field. To achieve these goals, a number of
plausible, nevertheless important assumptions were made, which are 
described next.
 
The galaxy distribution is assumed to be a discrete, locally
Poissonian realization of an underlying smooth random field. To 
estimate the moments of the PDF of this random field, 
factorial moments (e.g.~Szapudi \& Szalay 1993a) of the count 
probability distribution function $P_N(\ell)$ (CPDF) are used. 
By definition, the CPDF represents the probability of finding 
$N$ galaxies in a spherical (circular) cell of radius $\ell$ thrown at
random in a three-dimensional (two-dimensional) galaxy catalog. 
The CPDF is easy to measure and widely used to test the {\em
scaling behavior} of galaxy catalogs (see e.g.~Alimi, Blanchard \& 
Schaeffer 1990; Maurogordato, Schaeffer \& da Costa 1992; Szapudi,
Szalay \& Boch\`an 1992; Bouchet et al. 1993; Gazta\~naga 1992, 1994; 
Szapudi, Meiksin \& Nichol 1996, Szapudi \& Szalay 1997a)
and $N$-body simulations data sets (see e.g.~Bouchet, Schaeffer \& 
Davis 1991; Bouchet \& Hernquist 1992; Baugh, Gazta\~naga \&
Efstathiou 1995; Gazta\~naga \& Baugh 1995; Colombi, Bouchet \& 
Hernquist 1996).

Throughout this work, we assume an ideal three-dimensional 
magnitude limited galaxy catalog ${\cal E}$, of depth $R_{\rm max}$, 
with magnitude limit $M_{\rm lim}$, covering a given volume $V$ of the 
universe, and containing a number $N_{\rm obj}$ of spherical
coordinates of galaxies, $(z,\theta,\phi)$, $z$ being the measured 
redshift. 

A purely statistical approach is used: except for the effects of the 
selection function in a magnitude limited sample, all observational 
effects are ignored, such as extinction, confusion limit,  
and systematic errors due to the imperfection of the instruments.

Redshifts are considered as pure distances, i.e.~effects of projection
in redshift space are neglected. Such effects can significantly change
the behavior of the CPDF, especially in the nonlinear regime
(e.g.~Kaiser 1987; Lahav et al.~1993; Matsubara \& Suto 1994; Hivon et 
al.~1995).

Finally, it is assumed that the clustering of galaxies does not depend 
significantly on their luminosity, which is probably a crude
approximation. Indeed, there are both theoretical (e.g.~White et al. 
1987; Mo \& White 1996;  Valls-Gabaud, Alimi \& Blanchard 1989; 
Bernardeau \& Schaeffer 1992, hereafter BeS) and observational
arguments (e.g.~Hamilton 1988; Davis et al.~1988; Dominguez-Tenreiro \&
Martinez 1989; Benoist et al.~1996) suggesting that the level of
clustering of galaxies increases with their luminosity. 

In realistic redshift surveys, the average number density of
galaxies $n(r)$ changes with the distance $r$ from the observer.
The CPDF is traditionally defined in an {\em homogeneous} catalog, 
i.e.~with constant $n(r)$. One way to bypass this problem is 
extraction of volume-limited subsamples ${\cal E}_{\rm VL}^i$ of 
depth $R_i \leq R_{\rm max}$ from the main catalog ${\cal E}$ 
(see for example Maurogordato et al.~1992; Bouchet et al.~1993).
In these subsamples, the objects are such that their apparent 
magnitude at distance $r=R_i$ would be larger than $M_{\rm lim}$.
Such a selection criterion renders the number density of galaxies 
in the catalog independent of distance, at the price of significant 
information loss. Although this can be partially recovered by 
cutting several volume-limited catalogs with various values of $R_i$, 
it would be preferable to extract all the available information 
from the catalog in a single measurement. This is possible by defining an
inhomogeneous counts-in-cells measure which is corrected for the 
variation of $n(r)$ with the distance from the observer, 
similarly to, e.g., Efstathiou et al.~1990 and Szapudi \& Szalay 1996. 
This, combined with a minimum variance weighting, results in 
unbiased, selection corrected estimators of the $N$-th factorial
moment of the galaxy counts. To clarify the substantial gain from 
such an approach, we carefully calculate the errors on the
measurements, and show how our method minimizes them. 

Only statistical errors are considered, 
caused by the fact that only a finite part 
of the universe is accessible for observations. The 
corresponding {\em cosmic error} was calculated by Szapudi \& Colombi 
(1996, hereafter SC) for count-in-cells measurements in an homogeneous
catalog. The different contributions were classified as follows:
\begin{enumerate}
\item The finite volume error is due to fluctuations
of the underlying random field at wavelengths larger than the size of the
catalog. The finiteness of the sampled volume causes
systematic effects on the measurement of the CPDF and its moments,
even in $N$-body simulations (Colombi, Bouchet \& Schaeffer 
1994, 1995, hereafter CBSI and CBSII). 
\item Edge effects are related to the geometry of the catalog: 
the galaxies near the edges of the catalog receive less statistical 
weight than those far from the boundaries. Edge effects can be theoretically
corrected for, at least partially. A corrected estimator can be used for the
two-point correlation function $\xi(r)$ (see, e.g.~Ripley 1988, p.~22; 
Landy \& Szalay 1992) from which the variance of the PDF, ${\bar
\xi}(\ell)$, can be obtained as a double integral of $\xi$ over a cell 
of radius $\ell$. A class of edge corrected estimators was actually
recently introduced by Szapudi \& Szalay (1997b) for the higher
order moments as well. They, however, use a somewhat complicated procedure 
which is applicable only to the moments of fluctuations.
\item The shot noise error (or, equivalently, the error from
discreteness effects) is due to the incomplete sampling of the 
underlying smooth field with a finite number of points. In particular,
excessive undersampling causes degeneracy of the CPDF, rendering the 
confrontation of various models against observations difficult
(e.g.~Bouchet et al.~1993, CBSII). By definition, discreteness effects tend
to zero as the average number density of objects in the catalog
approaches infinity\footnote{Note however that part of the shot noise 
belongs to the subclass of ``edge-discreteness'' effects, i.e.~it is
also an edge effect, and can be corrected for with appropriate techniques
(Szapudi \& Szalay 1997b).}.
\end{enumerate}

While unbiased estimators can be constructed in a number of ways,
the above sources of errors can be reduced by giving an appropriate
statistical weight depending on the region in the catalog.
In a magnitude limited catalog for instance, density decreases
far away from the observer, resulting in an increase in
the shot noise. This alone would call for
increasing statistical weight close to the observer, 
thus decreasing the contribution of distant portion of the survey. 
However, this reduces the effective sampled volume, and 
increases the finite volume and the edge contributions to the error.
There exists a ``minimum variance'' weighting which provides
a compromise between various effects. By deriving and
solving the integral equation for the minimum variance weighting
scheme, we show that a significant gain in accuracy
on the measurement can be achieved with our method, compared to the traditional 
volume-limited approach. Note that the word ``optimal'' is used
synonymously with minimum variance in this paper, i.e. producing the
smallest error in a class of unbiased estimators. Although this usage
does not emphasize it, it is possible in principle to find even better
estimators by extending the class in which the search is performed.

While the above procedure can optimize the way information is retrieved
from an existing survey, calculation of cosmic errors helps in
designing optimal surveys. In particular, an optimal sampling
strategy can be found to build a three dimensional galaxy 
catalog for a given statistical indicator. Given the available telescope time,
K86 found the optimal
strategy for the measurement of the two-point correlation function of
galaxies, using a simple model 
for the cosmic error. To reduce finite volume and edge effects, 
which are independent of the number of objects in the catalog,
a sparse survey with large volume is needed. To reduce discreteness 
effects, on the other hand, a catalog as dense as possible is
preferable. These competing effects determine the optimal sampling
strategy with respect to a particular statistic. For the two-point
function, K86 concluded in favor of sparse samples with approximately 
$1/10$ to $1/20$ of the candidates randomly selected for measuring
their redshifts. Sparse sampling strategies were in fact used for
several surveys, such as the Stromlo APM redshift survey (e.g.~Loveday
et al.~1992), the QDOT redshift survey (e.g.~Moore et al.~1994), and
the Durham/UKST redshift survey (e.g.~Ratcliffe et al.~1997). However,
the advent of multi-fiber spectroscopy makes large, complete galaxy
surveys possible in a significantly shorter time than when K86
proposed his idea. Prime examples are the Sloan Digital Sky Survey
(SDSS, e.g.~Loveday 1996), and the 2DF Survey (e.g.~Lahav 1996). 
While the problem of
optimal sampling strategies might loose from its relevancy with these
new developments, it is still worth to study it: either to facilitate
preliminary investigations to prepare large surveys, or optimize the
design of surveys aimed at statistical properties of rarer or harder
to find objects at various wavelengths. Indeed, it will be shown how
sparse sampling strategies are relevant for measuring low order
moments in a reasonable scale range, while high order statistics,
direct analysis of the CPDF shape, cluster selection, etc., are 
highly sensitive to discreteness effects, 
therefore proving that full sampling is more advantageous.

Another important design issue, when the catalog has poor sky coverage, 
is its geometry (e.g.~Kaiser 1996, hereafter K96).  This affects 
finite volume, edge, and 
shot noise effects in different ways. Although a general solution to this
complicated problem will not be given, it will be discussed in broad
terms, with suggestions for reasonable design principles.
The solution for the conceptually simplest case will be given, which
is relevant even for surveys which will eventually cover
a continuous portion of the sky. The results presented here
can facilitate the extraction of preliminary results before
the catalog is finished. 

This article is organized as follows. \S~2 introduces the 
statistical indicators used along this paper, i.e. the weighted
factorial moments corrected for selection effects. \S~3 contains the
calculation of the general expression for the cosmic error, extending
the results of SC to a magnitude limited catalog with a spatial
weight. The integral equation for the optimal weight is given. Some
simple but still quite general assumptions
on the underlying statistics are made to enable the numerical
calculation of the optimal weight and the corresponding
cosmic error. \S~4 elaborates a practical example: a SDSS-like
catalog. In particular, a useful approximation for the
optimal weight is found. We emphasize the advantages of extracting the 
information from the data with the optimal weight, compared 
to the traditional volume limited approach. In \S~5, sparse
sampling strategies are studied along with
the suitable choice of  geometry for the catalog. \S~6
summarizes and discusses the results. In addition, three
Appendices provide further information.
\S~A contains mathematical details concerning 
the calculation of the cosmic error. \S~B explains and tests
the numerical method used to estimate the optimal weight. 
\S~C contains mathematical formulae used to derive analytical
expressions for the optimal sampling rate in some asymptotic regimes. 
%
%
\section{Weighted factorial moments corrected for selection effects}
%
%
For a statistically homogeneous catalog, the CPDF $P_N$ is
the probability of finding $N$ galaxies in a cell of size $\ell$.
The factorial moments of counts in cells are defined for $k \geq 0$ by
\begin{equation}
    F_k(\ell) \equiv \avg{(N)_k}  \equiv \
    \sum_N (N)_k P_N,
    \label{eq:fkdef}
\end{equation} 
where the falling factorials are defined as
$(N)_k = N (N-1) \ldots (N-k+1)$, and $(N)_0 \equiv 1$, 
therefore $F_0\equiv 1$. Under the assumption of infinitesimal
Poisson sampling (Peebles 1980, p.~147) 
the factorial moments correct directly for the discrete nature of 
galaxy catalogs. More precisely, their ensemble averages 
are equal to the moments of the (appropriately normalized) 
underlying {\em smooth} density field $\rho$ ($\avg{\rho} = 1$),
\begin{equation}
    F_k = {\bar N}^k \langle \rho^k \rangle, 
    \label{eq:fknbar}
\end{equation}
where $ {\bar N} \equiv \langle N \rangle = F_1$
is the average number of objects per cell
(see Szapudi \& Szalay 1993a).

Counts-in-cells statistics are estimated by a large number
of sampling cells positioned at random in the catalog.
For a homogeneous survey, the corresponding estimator 
of the factorial moment is expressed by the counts $N_i$ in
the $i$th cell, $1 \leq i \leq C$,
\begin{equation} 
    \tF_k^C \equiv \frac{1}{C} \sum_i^C \left( N_i \right)_k.
    \label{eq:form3}
\end{equation}
Equivalently, the CPDF can be estimated first (e.g.~SC), 
which gives the above estimator through equation~(\ref{eq:fkdef}).
The form (\ref{eq:form3}) is meaningful if all the cells are equivalent, 
i.e.~if they all have the same statistical weight $\omega=1$. 
In an attempt to correct for finite volume and edge effects, 
a weight $\omega_{\ell,k}(\r_i)$ (to be determined later)
can be assigned to each cell $i$. This can
depend on the position and size of the cell, and the statistic
at hand:
\begin{equation} 
    \tF_k^C \equiv \frac{1}{C} \sum_i^C \left( N_i \right)_k 
	\omega_{\ell,k}(\r_i).
    \label{eq:faci}
\end{equation}
The weight $\omega_{\ell,k}$ is determined by minimizing the
value of the cosmic error under the constraint of
appropriate normalization.

A magnitude limited catalog ${\cal E}$, such as defined in introduction, 
is inhomogeneous since the selection function is not uniform. 
If $n$ is the real number density of the underlying galaxy 
distribution (assuming that such number is well defined), 
the effective number density $n(r)$ of galaxies in a thin shell 
at distance $r$ from the observer reads
\begin{equation}
  n(r)=n \phi(r),
\end{equation}
where $\phi$ is the selection function. In particular, the average
number of galaxies in a cell of radius $\ell$ at distance $r$ from the
observer is given by
\begin{equation} 
 {\bar N}_r =  n_\ell(r) v, \quad v \equiv \frac{4}{3} \pi \ell^3,
	\label{eq:nbarn}
\end{equation}
with
\begin{equation}
  n_\ell(r)=n \phi_\ell(r),
\end{equation}
and $\phi_\ell(r)$ is the average of the selection function over a
cell. The approximation $\phi_\ell\simeq \phi$ is excellent
for the relevant scales, valid within a few percents at worst. 
It will be used for practical calculations throughout.

To correct for selection effects we follow Szapudi \& Szalay (1996) 
by changing equation (\ref{eq:faci}) in
\begin{equation}
  \tF_k^C \equiv \frac{1}{C} \sum_{i=1}^C \frac{\left( N_i \right)_k 
  \omega_{\ell,k}(\r_i)}{[\phi_\ell(r_i)]^k}.
  \label{eq:myw}
\end{equation}
This is the final form we propose for the estimator of the factorial
moments. Note that {\em a priori} knowledge of the 
selection function $\phi(r)$ is assumed. Although implicitly always
present, the $k$ and $\ell$ dependence of the weights
$\omega_{\ell,k}$ will be usually omitted. The normalization of the 
weights follows from the requirement that the estimator is unbiased. 
Taking the ensemble average of the above equation, and averaging
all possible random realizations of $C$ cells, gives
\begin{equation}
   \frac{1}{{\hat V}(\ell)}\int_{{\hat V}(\ell)} d^3r \omega(\r) 
   = 1,
  \label{eq:omnorm}
\end{equation} 
where ${\hat V}(\ell)$ is the {\em effective} sampled volume, \ie the
volume occupied by the center of all possible cells of radius $\ell$ 
{\em contained} in the catalog. While the proposed estimator can be
used directly to estimate the factorial moments, it might be
advantageous to introduce the inhomogeneous CPDF, $P_N(r)$. 
Once this is estimated, the inhomogeneous factorial moments $F_k(r)$
can be calculated. Using the scaling $\phi(r)^{-k}$ and summing over
with the appropriate weights $\omega(\r)$ is identical to the proposed estimator.
This way, however, any high order moment can be calculated using
the inhomogeneous CPDF without the need of rescanning the whole catalog. 
%
%
\section{Cosmic error and optimal  weight}
%
%
This section generalizes the formalism for computing the cosmic error
presented by SC to the case of inhomogeneous selection function and weight.
The equation for the optimal  weight is derived in \S~3.1. 
The locally Poissonian and hierarchical assumptions are used to
simplify the calculations in \S~3.2.
A few comments follow on the interpretation of the
results thus far (\S~3.3), and, finally, it is shown 
how this formalism can be applied to practical measurements (\S~3.4).
%
\subsection{Formalism}
%
The variance of $\tF_k^C$  is defined by 
\begin{equation}
  \left( \Delta \tF_k \right)^2 \equiv \left\langle \left\langle
  \tF_k^2 \right\rangle \right\rangle_C 
   - \left\langle \left\langle \tF_k \right\rangle \right\rangle_C^2,
  \label{eq:deltaFk}
\end{equation}
where $\avg{\ }$, and $\avg{\ }_C$ are the ensemble average, and averaging
over all possible sets of sampling cells, respectively (we dropped the $C$
dependence of $\tF_k$).
For the proposed estimator
\begin{eqnarray}
  \displaystyle
  \left\langle \left\langle \tF_k^2 \right\rangle \right\rangle_C &=&
   \displaystyle
   \left\langle \frac{1}{C^2} \sum_{i=1}^C \omega^2(\r_i)
   \frac{\left\langle (N_i)^2_k \right\rangle}{[\phi(r_i)]^{2k}} 
   \right\rangle_C  \nonumber\\
   &+& \displaystyle 
   \left\langle \frac{1}{C^2} \sum_{i\neq j}^C \omega(\r_i)\omega(\r_j) 
   \frac{\left\langle (N_i)_k(N_j)_k \right\rangle}
   {[\phi(r_i)\phi(r_j)]^k} \right\rangle_C.
   \label{eq:facerbas}
\end{eqnarray}
Following SC, the evaluation of this expression in terms
of the parameters of the distribution is facilitated by
generating functions. Let us introduce the generating function, $P_r(x)$,
of the probability $P_N(r)$ of finding $N$ objects in a cell of size
$\ell$ at position $r$, where the effective average number density is
$n_\ell(r)$:
\begin{equation}
   P_r(x) \equiv \sum_{N=0}^{\infty} x^N P_N(r).
   \label{eq:genr}
\end{equation}
Similarly, we define $P_{\r_i,\r_j}(x,y)$ as the generating function of
the bivariate probability $P_{N,M}(\r_i,\r_j)$ of finding $N$ and $M$
galaxies in cells of radius $\ell$ respectively at positions $\r_i$
and $\r_j$: 
\begin{equation}
   P_{\r_i,\r_j}(x,y) \equiv \sum_{N,M=0}^{\infty} x^N y^M P_{N,M}(\r_i,\r_j).
\end{equation}
As SC, we write formally
\begin{equation}
  \left( \Delta \tF_k \right)^2 = \left.
  \big{[} \frac{\partial}{\partial x} \big{]}^k 
  \big{[} \frac{\partial}{\partial y} \big{]}^k
   E^{C,V}(x+1,y+1) \right|_{x=y=0}.
  \label{eq:facer}
\end{equation}
The generating function of the total error, $E^{C,V}(x,y)$,
is asymptotically the sum of two generating functions
\begin{equation}
   E^{C,V}(x,y)=\left( 1 -\frac{1}{C} \right) E^{\infty,V}(x,y) 
 	+ E^{C,\infty}(x,y).
\end{equation}
Function $E^{\infty,V}(x,y)$ generates the errors for hypothetical
surveys with finite volume $V$ and infinite number of sampling cells:
\begin{eqnarray}
   E^{\infty,V}(x,y) & = 
   \displaystyle
   \frac{1}{{\hat V}^2}& \displaystyle \int_{{\hat V}} 
   d^3r_1 d^3r_2 \omega(\r_1) \omega(\r_2) 
   [\phi(r_1)\phi(r_2)]^{-k} \nonumber \\
   & & \displaystyle \left\{ P_{\r_1,\r_2}(x,y)
	-P_{r_1}(x)P_{r_2}(y) \right\}.
   \label{eq:ecosm}
\end{eqnarray}
Function $E^{C,\infty}(x,y)$ generates the errors due the finite
number of sampling cells used to do the measurement\footnote{If the
weight is homogeneous and if there are no selection effects
($\omega=\phi=1$), as in SC, this function does not depend on the
survey volume. In that case, it generates the errors for hypothetical
surveys with infinite volume ($V=\infty$) and finite number $C$ of
sampling cells. This explains the formal notation
``$E^{C,\infty}(x,y)$''.}:
\begin{eqnarray}
	\renewcommand{\baselinestretch}{1.0}
    E^{C,\infty}(x,y)  = & \displaystyle \frac{1}{C} & \displaystyle
   \left\{ \frac{1}{{\hat V}} \int_{{\hat V}} d^3r \omega^2(\r)
   [\phi(r)]^{-2k} P_r(xy)  \right.  \nonumber \\
   & & \displaystyle \mbox{}-  \frac{1}{{\hat V}^2}  \displaystyle \int_{\hat V} d^3r \omega(\r) 
   [\phi(r)]^{-k} P_r(x) \nonumber \\
   & &\displaystyle \left. \int_{\hat V} d^3r \omega(\r) [\phi(r)]^{-k} P_r(y) \right\}.
\end{eqnarray}
Note that the generating function defined above can {\em only} be
used for the error on the $k$-th order moment, i.e.~for each $k$
a slightly different generating function must be used. The reason
for this is the implicit $k$ dependence of the weight $\omega$
(the scaling with the selection function could be taken into account
simply with the substitution $x \rightarrow \phi x - \phi + 1$). 
Although it would be possible to define a single but
three variate generating function for all orders, it is
simpler and more practical to use the above definition.
Also, for the sake of completeness, we quoted both
the cosmic error $E^{\infty,V}(x,y)$, and the measurement error 
$E^{C,\infty}(x,y)$. The former is an inherent property of the survey,
while the latter is related to the finite number of sampling cells $C$
used to do the measurement. Since this contribution can be rendered
arbitrarily small with massive ``oversampling'', $C \rightarrow
\infty$, or with an algorithm corresponding to $C = \infty$ (Szapudi
1997) it will not be considered further (see SC for a discussion on
the ``number of statistically independent cells'', i.e. the number of
cells needed to extract all the relevant information from the
catalog). In what follows, the upper indices will be dropped from the
cosmic error for simplicity, $E = E^{C,V} \simeq E^{\infty,V}$.

The above formalism provides the framework to compute the optimal 
sampling weight, which minimizes the cosmic error on the measurement of
$\tF_k$. Taking into account the normalization of the
weights leads to the following Lagrangian
\begin{eqnarray}
  {\cal L}[\omega, \lambda]  & \equiv & 
  \displaystyle \big[ \frac{\partial}{\partial x}
  \big]^k 
  \big[ \frac{\partial}{\partial y} \big]^k 
  \left. E(x+1,y+1) \right|_{x=y=0} \nonumber \\
  & & \mbox{} + 2 \lambda \left\{ \frac{1}{\hat V} 
  \int_{\hat V} d^3r \omega(\r) -1 \right\}.
\end{eqnarray}
The optimal weight $\omega$ is thus the solution of the following integral
equation
\begin{eqnarray}
	\renewcommand{\baselinestretch}{1.0}
   \displaystyle \frac{1}{\hat V} \int_{\hat V} d^3 r \displaystyle
   \omega(\r) 
   [\phi(r)\phi(u)]^{-k} \big[ \frac{\partial}{\partial x} \big]^k 
   \big[ \frac{\partial}{\partial y} \big]^k \nonumber \\
   \left.\left\{ P_{\r,\u}(x+1,y+1)- P_{r}(x+1)P_{u}(y+1)
   \right\}\right|_{x=y=0} \nonumber \\ 
    \displaystyle
     \mbox{} + \lambda = 0.
   \label{eq:optiw}
\end{eqnarray}
The constant $\lambda$ is determined by the normalization 
(\ref{eq:omnorm}).

%
\subsection{Approximations}
%
To compute the optimal  weight for measuring $F_k$
and the corresponding cosmic error, generating functions
$P_{\r,\u}(x,y)$ and $P_r(x)$ are needed. More precisely, 
after partial differentiation of order $k$ at $x=y=0$ 
in equations (\ref{eq:facer}), the 
factors $F_{l,m}(\ell,\r,\u)$ are needed, which are defined as
\begin{equation}
   F_{l,m}(\ell,\r,\u)=\langle \left( N \right)_l \left( M \right)_m 
   \rangle
   \label{eq:facor}
\end{equation}
up to order $l+m=2k$. These quantities generalize 
the concept of factorial moments to bivariate distributions (Szapudi
et al.~1995). In equation (\ref{eq:facor}), the ensemble average is
taken over two cells of size $\ell$ at positions $\r$ and $\u$ in 
the catalog, which can possibly overlap. This especially complicates
the problem of finding the optimal weight for measuring $F_k$ and the
corresponding cosmic error.

Following SC, one can make reasonable assumptions about the 
underlying statistics to simplify the calculations considerably:
the hierarchical tree model, and the locally Poissonian
approximation. Then, as shown below, only the factorial moments
$F_l$, $l \leq 2k$ and the two point-function $\xi(r)$ are required 
{\em a priori} for the computation of the optimal weight and 
the corresponding cosmic error on $F_k$.
The calculations are detailed in Appendix A. We 
summarize here the important steps and hypotheses:

\begin{enumerate}
\item The integral (\ref{eq:ecosm}) is split into two parts, 
according to whether or not the two cells overlap. 
\item The calculation of the non-overlapping part requires the
knowledge of the bivariate count-in-cells generating function for
disjoint cells. Following SC, we simplify it by taking two particular
but still quite general cases of the hierarchical model (e.g.~Peebles
1980, p.~206 \& 211, Balian \& Schaeffer 1989, hereafter BS) by SS and BeS 
(see Appendix A for more details). The hierarchical model is 
seen to be a good approximation for the higher order statistics in the
observed galaxy distribution (e.g.~Groth \& Peebles 1977; Fry \&
Peebles 1978; Sharp, Bonometto \& Lucchin 1984; Szapudi et al.~1992;
Meiksin, Szapudi \& Szalay 1992; Szapudi et al.~1995; Szapudi \&
Szalay 1997a) and in $N$-body simulations (e.g.~Efstathiou et al.~1988;
Bouchet et al.~1991; Bouchet \& Hernquist 1992; Fry, Melott \&
Shandarin 1993; Bromley 1994; Lucchin et al.~1994; CBSI, CBSII; 
Colombi et al.~1996).

The function $P_{\r_1,\r_2}(x,y)$ is then Taylor-expanded to first
order in $\xi(r_{12})/{\bar \xi}(\ell)$, where $\xi(r)$ is the
two-point correlation function, $r_{12}\equiv |\r_1-\r_2|$ and 
\begin{equation}
 {\bar \xi}(\ell)\equiv \frac{1}{v^2} \int_v d^3 r_1 d^3 r_2
 \xi(r_{12}).
\end{equation}
This approximation is becoming more accurate
when the cells are far away from each other,  however, it is
still reasonable even when the cells touch each other (e.g.~Bernardeau 1996). 
\item To compute the overlapping contribution, the variations of the
weight and of the selection function are neglected within the
cells. We also assume local Poisson behavior, which considerably
simplifies the writing of the bivariate generating function of counts
for overlapping cells (see SC). 
\end{enumerate}

SC carried out {\em explicitly} the calculation to leading order 
in $v/V$ for uniform survey and constant weight. They found that the 
cosmic error could be separated into three contributions
\begin{equation}
    \Delta^{2}[\omega=1,\phi=1]  \equiv 
    \displaystyle \left( \frac{\Delta \tF_k}{F_k}
    \right)^2 
    = \Delta^2_{\rm F}
    + \Delta^2_{\rm E} + \Delta^2_{\rm D}.
   \label{eq:cosmicout}
\end{equation}

The term $\Delta^2_{\rm F}$ is the {\em finite volume error} discussed in
introduction, arising from the contribution of disjoint cells in
equation (\ref{eq:ecosm}). It is proportional to the integral of the
two-point function over the sampled volume
\begin{equation}
  \xiav(L)\equiv \frac{1}{V^2} \int_V d^3r_1 d^3r_2 \xi(r_{12})
\end{equation}
($L \sim V^{1/3}$). While it depends on the clustering properties of the
system, i.e., 
\begin{equation}
   \Delta^2_{\rm F} = {\cal F}_{\rm F} 
   \left\{ F_l/{\bar N}^l, l \leq 2k \right\}\ \xiav(L),
\end{equation}
it is independent of the average number density $n$ [see
eq.~(\ref{eq:fknbar})]. Even if
the number of objects in the catalog increases, the
finite volume error remains {\em unchanged}. 

Similarly, the {\em edge error} $\Delta^2_{\rm E}$ formally written as
\begin{equation}
\Delta^2_{\rm E} = {\cal F}_{\rm E}\left\{ F_l/{\bar N}^l, l
  \leq 2k \right\}\ \xiav v/V,
\end{equation}
is independent of average number density as well. Edge effects are
caused by uneven weighting near the edges of the catalogue, thus
increasing with $v/V$.

Finally, the {\em discreteness (shot noise) error} $\Delta^2_{\rm D}$
can formally be written as
\begin{equation}
   \Delta^2_{\rm D}={\cal F}_{\rm D}\left\{ F_l, l
  \leq 2k \right\}\ v/V.
\end{equation}
This is the only contribution to the cosmic error 
which depends on average number density $n$. 
To leading order in ${\bar N}$, it is proportional to ${\bar
N}^{-k}v/V$ thus becomes important at small scales or when the number
of objects $N_{\rm obj}$ in the catalog is small.

SC have tested successfully the validity of the above approximations 
by computing the cosmic error in artificial galaxy catalogs generated
by Rayleigh-L\'evy random walks, for which 
the clustering properties could be calculated exactly (e.g.~CBSII). 

When $\omega \neq 1$ and $\phi(r) \leq 1$, equation 
(\ref{eq:cosmicout}) generalizes to (see Appendix A) 
\begin{equation}
   \Delta^2_{\rm cosmic}[\omega,\phi]  
     \simeq \Delta^2_{\rm F}[\omega]
     + \Delta^2_{\rm E}[\omega] + \Delta^2_{\rm
     D}[\omega,\phi],
   \label{eq:cosmic2}
\end{equation}
with
\begin{equation}
\Delta^2_{\rm F}[\omega]=\frac{\Delta^2_{\rm F}}
   {\xiav(L) {\hat V}^2} \int_{\hat V} 
   d^3r_1 d^3r_2 \omega(\r_1)\omega(\r_2) 
   \xi(r_{12}),
   \label{eq:delfinom}
\end{equation}
\begin{equation}
\Delta^2_{\rm E}[\omega] = \frac{\Delta^2_{\rm E}}{\hat V}
   \int_{\hat V} d^3r \omega^2(\r),
  \label{eq:deledge}
\end{equation}
\begin{equation}
  \Delta^2_{\rm D}[\omega,\phi]=
   \frac{1}{\hat V} \int_{\hat V} d^3r \omega^2(\r)  
    \Delta^2_{\rm D}(r).
   \label{eq:deldis}
\end{equation}
The $r$ dependence of $\Delta^2_{\rm D}$ in the integral 
(\ref{eq:deldis}) is caused by the selection function, as
this type of error depends on the average count.
On the other hand, the selection effects naturally 
canceled out in the finite volume and the edge errors,
as expected. 

The integral equation for the  optimal weight becomes
\begin{eqnarray}
   \displaystyle
   \frac{\Delta^2_{\rm F}} {\xiav(L) {\hat V}} 
   \int_{\hat V} d^3u \omega(\u)
   \xi(|\r-\u|)
    + \left\{ \Delta^2_{\rm E} +
    \Delta^2_{\rm D}(r)
    \right\} \omega(\r)  \nonumber \\
    \mbox{} + \lambda = 0,
   \label{eq:optiw2}
\end{eqnarray}
and the constant $\lambda$ is determined by the normalization
(\ref{eq:omnorm}). This standard equation  can
be solved numerically. 
%
\subsection{Interpretation}
%
From equation (\ref{eq:optiw2}), the following immediate conclusions
can be drawn on how different contributions to the cosmic error
influence the optimal weight: 
\begin{enumerate}
\item If edge effects were dominant, the optimal weight 
would simply be uniform. This is contrary to intuition 
suggesting  increasing weight at the edges to compensate for 
the lesser statistical weight carried by these regions of the catalog.
This is how edge effects are corrected for
the two-point function $\xi(r)$ (e.g.~Ripley 1988, p.~22).
The finite extension of the cells, however, prevents us from
correcting for edge effects with our indicator (\ref{eq:myw}), which
uses a simple multiplicative sampling weight. More involved additive 
correction for the moments of the {\it fluctuations} of counts in
cells will be explained elsewhere using the formalism outlined in
Szapudi \& Szalay 1997b.
\item If discreteness effects were dominant, the correct weight
approximately would be
\begin{equation} 
  \omega(\r) \propto 1/\Delta^2_{\rm D}(r),
\end{equation}
in agreement with intuition. Indeed, the effective number density
$n(r)$ decreases with increasing distance from the observer, thus the
shot noise error increases as well. This is compensated by
$\omega$ decreasing with $r$. 
\item Finite volume effects are difficult to predict without explicit
calculations. The correlation function $\xi(r)$ is expected
to follow approximately a power-law behavior
from observations 
(Totsuji \& Kihara 1969; Peebles 1974; Davis \& Peebles 1983)
\begin{equation} 
  \xi(r)=\left( \frac{r}{r_0} \right)^{-\gamma},\quad r_0 \simeq 5\
  h^{-1}\ {\rm Mpc}, \quad \gamma \simeq 1.8,
  \label{eq:xipow}
\end{equation}
for $0.1\ h^{-1}\ {\rm  Mpc} \la r \la 10\ h^{-1}\ {\rm Mpc}$. At larger
scales it decreases rapidly
with scale, becoming negative at scales around $\sim 30-100\ h^{-1}\
{\rm Mpc}$ (e.g.~Fisher et al.~1994; Tucker et al.~1996), although
this turnaround value is presently uncertain. After that, it is expected
to oscillate slowly around zero with very small amplitude. With such a
behavior, it is not obvious to predict the optimal weight 
without explicit numerical calculations. 
The result will depend on the size and the geometry
of the catalog. The only simplification is that if the finite volume
error were dominant, the corresponding optimal weight would be
independent of the order $k$.
\end{enumerate}
%
\subsection{Practical Measurements}
%
As shown in \S~3.2, even though it was considerably simplified with
reasonable hypotheses, the calculation of the optimal 
weight and the cosmic error for $F_k$ needs prior knowledge of $F_l$, 
$l \leq 2k$ (including $F_k$ itself !) and of $\xi(r)$ (and of course
of the selection function). 
%
%
%
%
As a result, we propose two procedures to perform a practical
measurement in a galaxy catalog:
\begin{enumerate} 
\item One possibility is to choose a model of large
scale structure with given values of $F_l$, $l \leq
2k$ and $\xi(r)$. These values are used as input parameters 
in equation (\ref{eq:optiw2}) to compute the optimal 
weight and the corresponding theoretical cosmic error
[eq.~(\ref{eq:cosmic2})]. The value of $F_k$ measured
with this weight can be compared to the theoretical one, 
given the theoretical cosmic error. This procedure should be applied
again to each competing model.
\item An alternative iterative approach starts with measuring 
directly the values of $F_l$, $l\leq 2k$ in the galaxy catalog, with a
given weight, for example $\omega=1$. With these values of
$F_l$, one would solve equation (\ref{eq:optiw2}) to find the optimal
weight for measuring $F_k$, and thus perform a more accurate
measurement. This can be repeated until convergence is achieved. We
conjecture that a small number of iterations should be sufficient in
practice. The main weakness of this model independent approach is that
there is a cosmic error on the determination of the optimal weight
itself. As discussed in SC (and earlier by CBSI and CBSII), the cosmic
error is likely to be {\em systematic}, which implies that the optimal
weight estimated this way might be {\em biased}. This bias, of course,
would only increase the cosmic error on the measurement of $F_k$: 
the estimator for the moments is still unbiased by definition. 
\end{enumerate}
%
%
\section{Example: the SDSS Catalog}
%
%
The future SDSS  will be a likely proving ground of the
methods proposed in this paper. Therefore a ``SDSS-like'' catalog,
${\cal E}$, is used to illustrate the applicability of the theory
outlined so far. This section is organized as follows. \S~4.1 presents
the properties of the hypothetical survey considered, i.e., its
luminosity function, geometry, and the underlying statistics
(function $\xi(r)$, factorial moments $F_k$). Then, in \S~4.2, the
optimal weight is computed with the corresponding cosmic error. A
simple approximation for $\omega(\r)$ is found which practically
minimizes the cosmic error and avoids solving numerically integral
equation (\ref{eq:optiw2}). Finally, in \S~4.3, we show the advantages
of our optimal strategy of using the full catalog compared to the
alternative of extracting volume limited subsamples. Details on the
method used to solve numerically equation (\ref{eq:optiw2}) are given
in Appendix B.

\subsection{Properties of the Survey}
%
%
The luminosity function of the catalog ${\cal E}$ is assumed to be of the
Schechter form (Schechter 1976)
\begin{equation}
  \varphi(L/L_*)=\phi_{*} (L/L_*)^{\alpha} \exp(-L/L_*),
  \label{eq:lumif}
\end{equation} 
with parameters taken from Efstathiou, Ellis \& Peterson (1988, see
also Efstathiou 1996)
\begin{equation}
  \alpha=-1.07, \quad \phi_*=0.0156 h^{3}\ {\rm  Mpc}^{-3},
\end{equation}
where $h$ represents the uncertainty of a factor two on the Hubble
constant: $H_0=100\ h$ km/s/Mpc. In what follows, we shall use
$h=0.5$, \ie
\begin{equation}
  H_0=50\ {\rm km/s/Mpc}.
\end{equation}
However, the results derived hereafter should not depend significantly
on the value of $H_0$. In fact, the Hubble constant influences the
shape of the power-spectrum of initial fluctuations through the
parameter $\Gamma=\Omega h^2$, where $\Omega$ is the density parameter
of the universe (e.g.~Efstathiou, Bond \& White 1992).

The average number density in a thin shell at distance $r$ from the 
observer is
\begin{equation}
  n(r)=\phi_* \Gamma\left[ \alpha+1,L_{\rm lim}(r)/L_* \right],
\end{equation}
where $L_{\rm lim}(r)$ is the minimum required luminosity for a galaxy
at distance $r$ from the observer to be included in the catalog, and
$\Gamma$ is the incomplete gamma function. If $K$-correction is
neglected,
\begin{equation}
   L_{\rm lim}(r)/L_*=10^{0.4(M_*-M_{\rm lim}+ 25 + 5 \log_{10}r)}.
\end{equation}
In the above equation, $r$ is expressed in Mpc. Our choice of $M_*$
and the magnitude limit $M_{\rm lim}$ is
\begin{equation}
  M_*=-19.68+5 \log(H_0/100), \quad M_{\rm lim}=18.3.
\end{equation}
Note that, with the above value of $\alpha$, $n(r)$ diverges
at $r=0$, formally implying that the average number density of the
real, total galaxy distribution is infinite. This, however, does not
affect the following calculations. 

For the geometry, we assume that the catalog covers a cone of depth
\begin{equation}
   R_{\rm max}=1200\ {\rm Mpc}
\end{equation}
with elliptic basis defined as follows in Cartesian coordinates:
\begin{equation}
  z^2 = \left( \frac{x}{\tan \theta_{\rm M}} \right)^2 +
   \left( \frac{y}{\tan \theta_{\rm m}} \right)^2,
\end{equation}
\begin{equation}
    \theta_{\rm M}=65^{\circ}, \quad \theta_{\rm m}=55^{\circ}.
\end{equation}
With the above choice of the parameters the catalog would typically
contain
\begin{equation}
   N_{\rm obj} \simeq 830000
\end{equation}
objects.

The underlying statistics is chosen as follows:
\begin{enumerate}
\item The two-point correlation function is an estimate of the nonlinear
matter autocorrelation function for the standard Cold Dark Matter
(CDM) model, which actually
contains the $H_0$ dependence of the subsequent calculations. 
It is computed by Fourier transform of the power-spectrum obtained
from the nonlinear ansatz of Peacock \& Dodds (1994). The
normalization is chosen such that the variance in a sphere of radius
$8\ h^{-1}$ Mpc is unity. The choice of the two-point function fixes
its average over a cell ${\bar \xi}$ and therefore the factorial
moment of order 2.
\item As the optimal weights for the measurement of $F_k$, $k \leq 4$
will be discussed, prior knowledge of the statistics up to $k=8$
is needed. Factorial moments $F_k$, $3 \leq k \leq 8$ are derived from
the measurements of Gazta\~naga (1994) of the coefficients $Q_N$:
\begin{eqnarray}
  Q_3= 1.35, \quad Q_4 =2.33, \quad Q_5=4.02,\nonumber \\ 
  \quad Q_6=6.7, \quad Q_7=10, \quad Q_8=12.
	\label{eq:qns}
\end{eqnarray}
By definition,
\begin{equation}
  Q_N \equiv \frac{1}{\Gamma_N v^N} \int_v d^3r_1 \ldots d^3r_N
  \xi_N(\r_1,\ldots,\r_N),
\end{equation}
where $\xi_N$ is the $N$-point correlation function (see e.g.~Peebles 
1980, p.~138; BS) and 
\begin{equation}
  \Gamma_N \equiv \frac{N^{N-2} {\bar N}^N {\bar \xi}^{N-1}}{N!}.
	\label{eq:gammaN}
\end{equation}
In general, $Q_N$ can depend on scale. We assume that the
hierarchical model applies, which implies that $Q_N =$ constant
independent of
scale. However, the cosmic errors should be fairly robust against small
variations of the cumulants.

The generating function of $P_N$ [eq.~(\ref{eq:genr})] 
can be expressed as
\begin{equation}
   P(x)=\exp\left[ \sum_{N=1}^{\infty} (x-1)^N \Gamma_N Q_N \right].
   \label{eq:genpn}
\end{equation}
This is a completely general equation, true outside of
the hierarchical model framework as well
(BS; Szapudi \& Szalay 1993a).

The factorial moments $F_k$ can be computed from the coefficients
$Q_N$, ${\bar N}$ and ${\bar \xi}$ through $F_k = [\partial/\partial
x]^k P(x)|_{x=1}$ [see eq.~(\ref{eq:fkdef})].
\end{enumerate}

The volume limited subsamples ${\cal E}^i_{\rm VL}$ 
extracted from our virtual catalog 
${\cal E}$ are of depths $R_i=200$, 400, 600, 800, 1000 and 1200 Mpc.
Typically, they are expected to contain respectively $N_{{\rm
obj},i}\simeq40100$, $138\ 500$, $208\ 800$, $211\ 800$, $164\ 900$ and
$104\ 200$ objects. 
%
%
\subsection{The optimal weight and the corresponding cosmic error}
%
The assumed catalog has a non-spherical geometry similar to the future
SDSS. Therefore the weight should depend both on the angles and
the distance from the observer $r$, except
when the finite volume error contribution is negligible. For simplicity,
however, only the radial direction will be considered, i.e.
 the cosmic error will be minimized 
in the subspace of functions $\omega(r)$.
After integration over the angles, the cosmic error becomes
\begin{equation}
   \Delta^2_{\rm cosmic}[\omega,\phi] \simeq
   \Delta^2_{\rm F}[\omega]+\Delta^2_{\rm
   E}[\omega]+\Delta^2_{\rm D}[\omega,\phi],
	\label{eq:newerr}
\end{equation}
with
\begin{eqnarray}
   \Delta^2_{\rm F}[\omega] & = & \displaystyle
   \frac{\Delta^2_{\rm F}}{\xiav(L) {\hat V}^2}
	\int_{{\hat R}_{\rm min}}^{{\hat R}_{\rm max}}
	\int_{{\hat R}_{\rm min}}^{{\hat R}_{\rm max}} 
        r_1^2 dr_1 r_2^2 dr_2 \nonumber \\
        & & \displaystyle {\hat \Omega}(r_1) \omega(r_1)
	{\hat \Omega}(r_2)\omega(r_2) {\tilde \xi}(r_1,r_2),
	\label{eq:newerr1}
\end{eqnarray}
\begin{equation}
   \Delta^2_{\rm E}[\omega] = 
   \frac{\Delta^2_{\rm E}}{\hat V} 
	\int_{{\hat R}_{\rm min}}^{{\hat R}_{\rm max}}
	r^2 dr {\hat \Omega}(r) \omega^2(r),
	\label{eq:newerr2}
\end{equation}
\begin{equation}
   \Delta^2_{\rm D}[\omega,\phi]=
	\frac{1}{\hat V} \int_{{\hat R}_{\rm min}}^{{\hat R}_{\rm max}}
	r^2 dr {\hat \Omega}(r) \omega^2(r) \Delta^2_{\rm D}(r),
	\label{eq:newerr3}
\end{equation}
where ${\hat \Omega}(r)$ is the solid angle covered by cells at
distance $r$ from the observer, ${\hat R}_{\rm
min}$, ${\hat R}_{\rm max}$ denote the distance of the closest 
and furthest cell to the observer, and
\begin{eqnarray}
	{\tilde \xi}(r_1,r_2) & \equiv & \displaystyle
	\frac{1}{{\hat \Omega}(r_1) {\hat \Omega}(r_2)}
	\int_{{\hat \Omega}(r_1)} \int_{{\hat \Omega}(r_2)} \nonumber \\
	& & \quad \sin\theta_1 d\theta_1 d\varphi_1
	\sin\theta_2 d\theta_2 d\varphi_2 
        \xi\big( [r_1^2 + r_2^2  \nonumber \\
        & & \quad \mbox{} - 2 r_1 r_2 
	(\cos(\varphi_1-\varphi_2) \sin\theta_1 \sin\theta_2 \nonumber
	\\
        & & \quad  \mbox{} +\cos\theta_1 \cos\theta_2) ]^{1/2}
	\big).
	\label{eq:defxit}
\end{eqnarray}
The optimal radial weight
is the solution of the following integral equation (provided
that a solution exists), 
\begin{eqnarray}
    	\frac{\Delta^2_{\rm F}}{\xiav(L) {\hat V}}
	\int_{{\hat R}_{\rm min}}^{{\hat R}_{\rm max}}
	u^2 du {\hat \Omega}(u) \omega(u) {\tilde \xi}(r,u)
	\nonumber \\ \displaystyle
        +\left\{ \Delta^2_{\rm E}+\Delta^2_{\rm D}(r)
	\right\} \omega(r)  + \lambda=0.
	\label{eq:optiwsph}
\end{eqnarray}

\begin{figure}
\centerline{\hbox{\psfig{figure=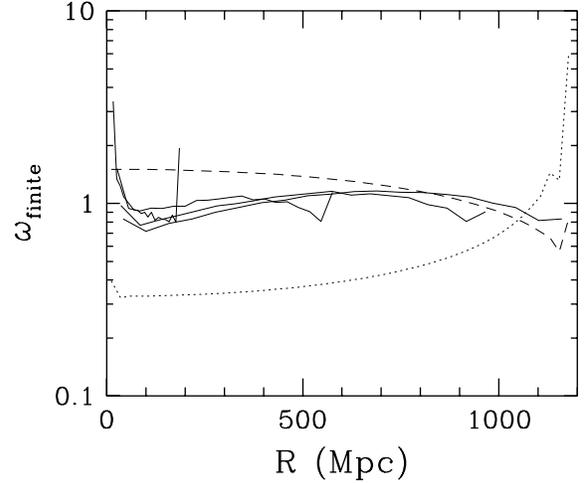,width=8cm}}}
\caption[]{The optimal weight $\omega$ for measuring the
factorial moment $F_k(\ell)$ ($k \geq 1$) is plotted as a function of
distance $R$ from the observer in the case the finite volume error is
dominant. The results displayed here correspond to $\ell=10$ Mpc;
other values of $\ell$ are similar. 
The solid curves correspond to various volume limited
subsamples ${\cal E}^i_{\rm VL}$ extracted from our SDSS like 
catalog ${\cal E}$ (see \S~4.1). The depth $R_i$ of the subsamples
increases with the $x$-coordinate of the right end point of the
curves: respectively $R_i=200$, 600, 1000 and 1200 Mpc. The
latter case is valid for the parent sample ${\cal E}$ as well. The
dashed curve corresponds to a catalog exactly the same as ${\cal E}$
but covering the full sky. The dotted curve is the same, but
the two-point function is assumed to be a power-law 
${\bar \xi}(\ell)=(\ell/18)^{-1.8}$ over all the available dynamic
range.}
\label{fig:figure1}
\end{figure}

Figure~\ref{fig:figure1} shows the optimal weight given by the
numerical solution of equation (\ref{eq:optiwsph}) in various
situations (see Appendix B for details of the numerical method).
The finite volume error contribution is assumed to be dominant
for this plot. In this case, the optimal weight is independent of the
statistical object under study, i.e., for $F_k$, of the order $k$. The
long dashes correspond to a catalog with similar characteristics as
our SDSS-like catalog, but covering the full sky, in order to have
the exact solution $\omega(\r)$ only depending on $r$. 
The dots correspond to the same
situation, but the two-point correlation function is assumed to be
${\bar \xi}(\ell)=(\ell/16)^{-1.8}$ over all the available dynamic
range, with $\ell$ expressed in Mpc. 
The four solid curves
correspond to four volume limited catalogs ${\cal E}^i_{\rm VL}$ of
our SDSS-like survey, of depths $R_i=200$, 600, and 1000 and 1200
Mpc. The case $R_i = 1200$ Mpc is valid as well for the parent
sample ${\cal E}$. %
The discontinuities on the extremities of the curves are boundary
effects related to the way we discretize the integral equation
(\ref{eq:optiwsph}). However, except for the left extremity of the
dotted curve, they are likely to express the fact numerically that the
optimal weight from finite volume effects is singular at the edges of
the survey, at least for the right extremity of the dotted curve (see
Appendix B.2).
There are also some 
small irregularities on the solid curves, but these are random 
fluctuations due to the finite number of steps in the Monte-Carlo simulation 
used for computing the angular average (\ref{eq:defxit})
(see Appendixes B.1, B.2). 
Apart from these details, we see that the optimal weight is a smooth
function, close to unity. Actually, taking $\omega=1$ gives
almost the same value as the optimal weight for the finite volume
error, at least for all the examples considered here. Thus a
homogeneous weight $\omega=1$ approximately minimizes the finite
volume error in the space of radial weights $\omega(r)$. Rigorously,
this result is not necessarily true for a catalog with a complicated
geometry. However, it should be valid if the catalog is compact
enough, which will be assumed in the following.

As a result, we propose the following approximation for the optimal
 weight, in the general case:
\begin{equation}
	\omega(r) \propto 1\left/ \left[ \Delta^2_{\rm F}
	+\Delta^2_{\rm E}+\Delta^2_{\rm D}(r) \right] \right..
	\label{eq:goodapp}
\end{equation}
The coefficient of proportionality is determined by the
normalization (\ref{eq:omnorm}). This approximation is quite natural,
because it properly takes into account the relative weight of each
contribution to the cosmic error. If one of them is dominant, then
$\omega$ converges to the correct solution of the integral equation 
(\ref{eq:optiw2}), although only approximately if the finite
volume error is dominant. 

\begin{figure*}
\centerline{\hbox{\psfig{figure=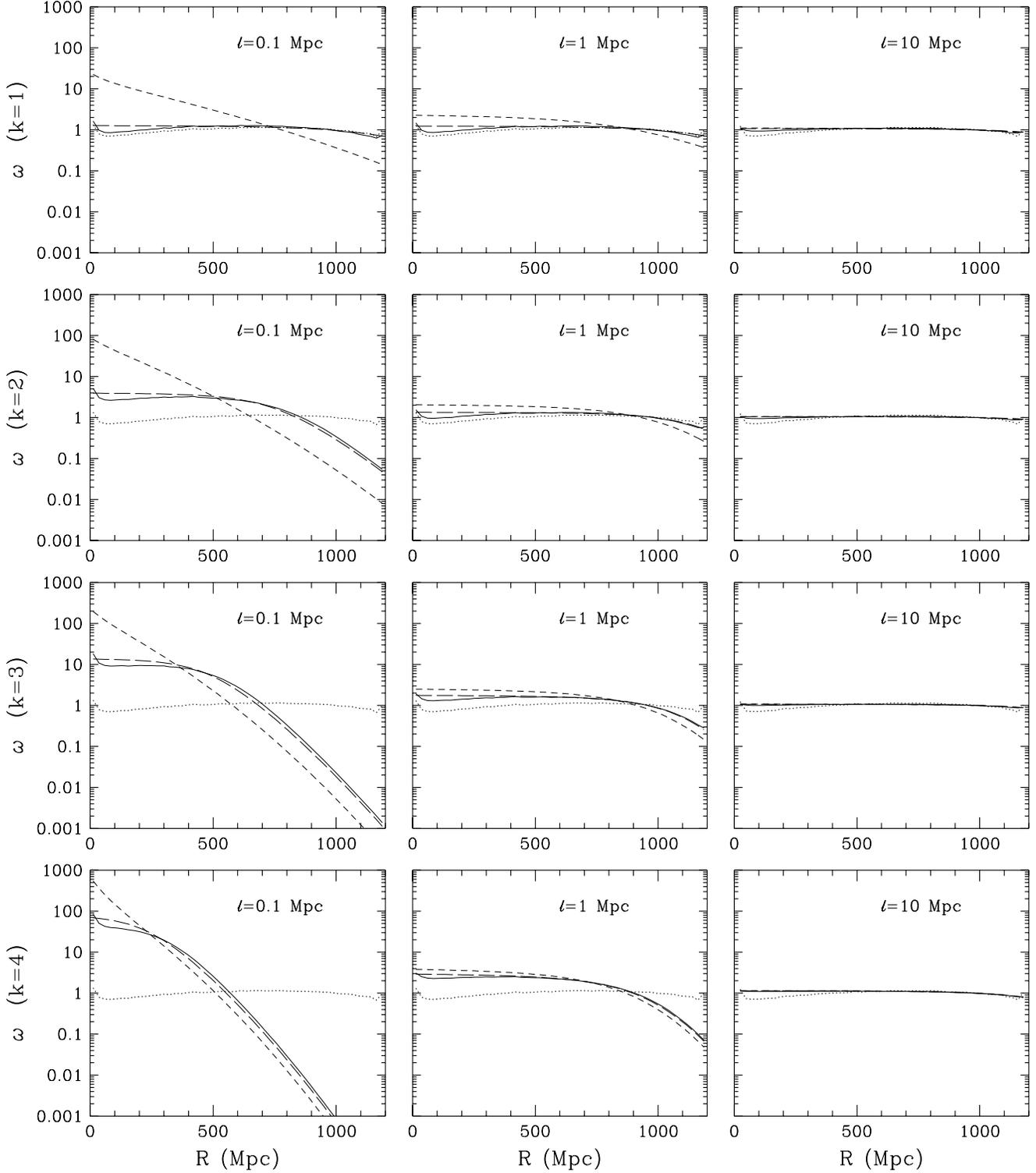,height=20.5cm}}}
\caption[]{The optimal  weight for measuring $F_k$ in our
virtual SDSS like catalog (see \S~4.1) is plotted as a function of the distance
$R$ from the observer (solid curves). Each panel corresponds to a
choice of $(k,\ell)$. The order $k$ increases from top to bottom and
the scale from left to right. The dashed curves are in the
assumption that the finite
volume error is negligible, while the reverse is true for the dotted
curves. The long-dashed curves correspond to our proposed approximation
(\ref{eq:goodapp}) for the optimal weight.}
\label{fig:figure2}
\end{figure*}

Figure~\ref{fig:figure2} shows the optimal (radial)  weight
obtained from the numerical solution of the integral equation
(\ref{eq:optiwsph}) (solid curve on each panel). 
Each line of panels corresponds to a given value of
$k$, which increases from top to bottom. Each column of panels
corresponds to a fixed value of the scale $\ell$ (from left to right,
$\ell=0.1$ Mpc, $1$ Mpc and $10$ Mpc). The dashes and dots display the
case when the  finite volume error is negligible,
and dominant, respectively.
The long dashes show approximation (\ref{eq:goodapp}).
They overlap surprisingly well with the solid curves. 

\begin{figure}
\centerline{\hbox{\psfig{figure=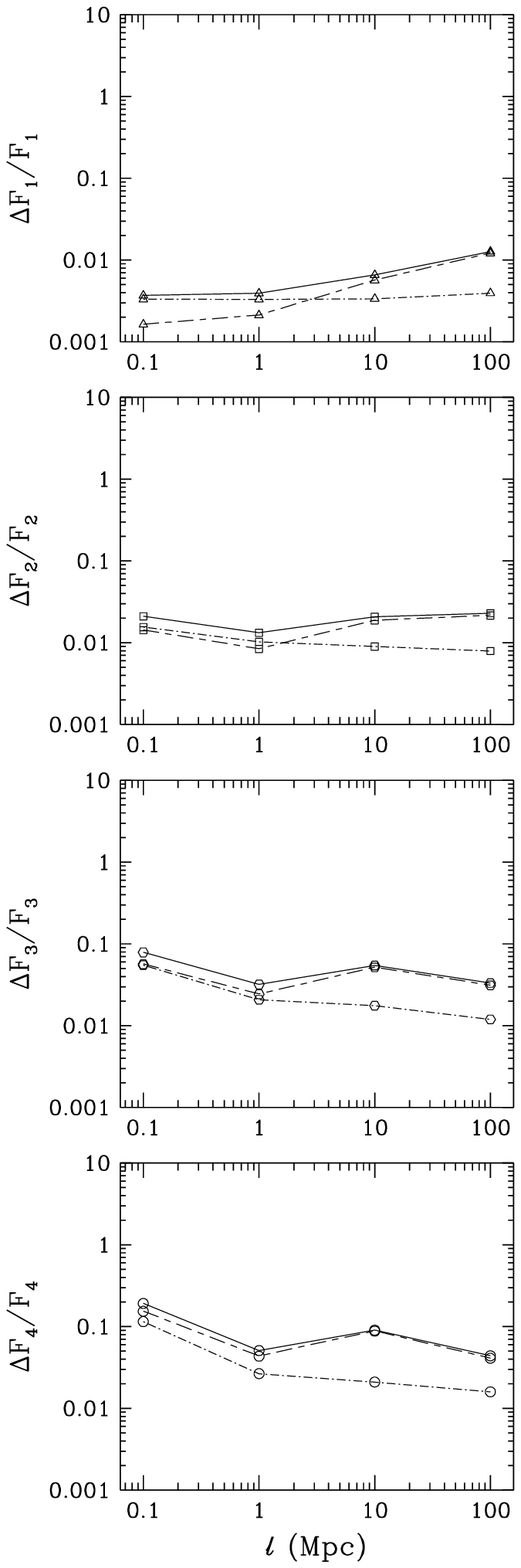,height=21.2cm}}}
\caption[]{The cosmic error is shown as a function of scale, when the
factorial moments $F_k$ are measured with the optimal  weight in
our virtual SDSS like catalog ${\cal E}$ (see \S~4.1). Each panel
corresponds to a value of the order $k$. The solid, dotted-dashed and
long-dashed short-dashed curves correspond respectively
to the total error, the finite volume, and
the edge plus discreteness effect contribution.}
\label{fig:figure3}
\end{figure}

Note that at large scales, the optimal weight tends to unity, because
discreteness effects become negligible, and edge effects dominant
(e.g.~SC). (The case of $\ell=100$ Mpc is not shown, since 
it is quite similar to $\ell=10$ Mpc). The departure of $\omega$ from
unity on the other hand increases at higher order $k$, and smaller
scales. Then discreteness effects tend to dominate the cosmic error
(e.g.~SC), and, as discussed in \S~3.3, the weight strongly
(exponentially) decreases with $r$. These arguments are partly
illustrated by Figure~\ref{fig:figure3}. For each value of $k$, various
contributions to the cosmic error calculated from equations
(\ref{eq:newerr}), (\ref{eq:newerr1}), (\ref{eq:newerr2}) and 
(\ref{eq:newerr3}), using the optimal weight, are plotted
as functions of scale. The solid, dotted-dashed, and long dashed-short
dashed lines correspond respectively to the cosmic error, the finite
volume error, and the edge effect plus shot noise contribution.

\begin{figure}
\centerline{\hbox{\psfig{figure=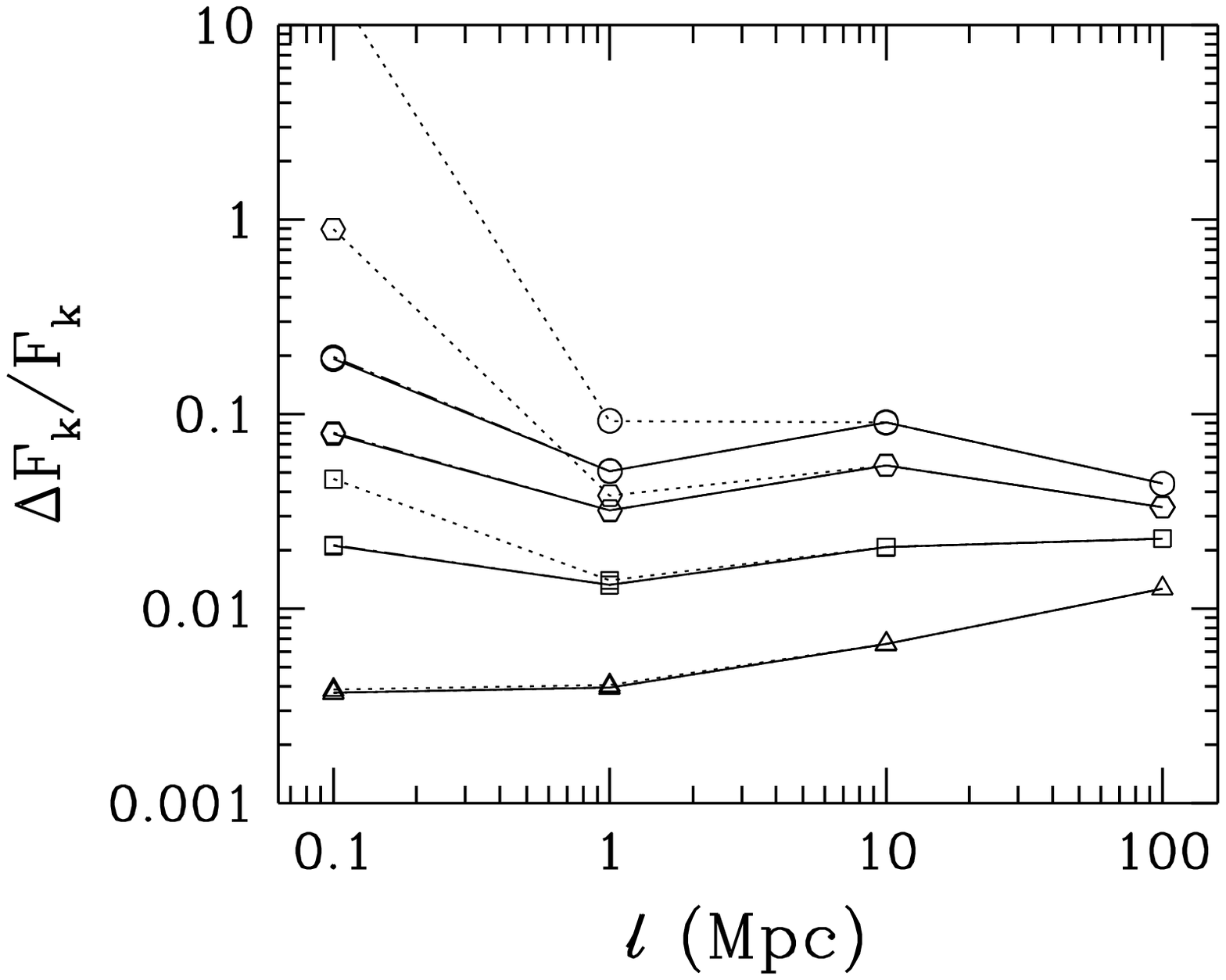,width=8cm}}}
\caption[]{The cosmic error for our virtual SDSS like catalog ${\cal
E}$ (see \S~4.1) is shown as a function of scale, when 
the factorial moments $F_k$ are measured with the optimal  weight
(solid curves), approximation (\ref{eq:goodapp}) (dotted-dashed
curves almost perfectly matching the continuous lines), and 
uniform weights (dots). The triangles, squares, hexagons and
circles respectively correspond to $k=1$, 2, 3 and 4.}
\label{fig:figure4}
\end{figure}

In figure \ref{fig:figure4}, the cosmic error is displayed as a
function of scale for all values of $k$. The solid lines
correspond to the result given by the optimal  weight and the
dots to $\omega=1$. There are also dotted-dashed
lines almost perfectly matching the solid ones: they correspond
to approximation (\ref{eq:goodapp}). The degree of matching
suggests that this is indeed an excellent approximation.
Triangles, squares, hexagons and circles respectively
correspond to $k=1$, 2, 3 and 4: the cosmic error increases with the
order $k$. According to the figure, 
$\omega=1$ provides a satisfactory weighting scheme
for our mock SDSS catalog on scales larger than $\sim 1$ Mpc.
%
%
%
%
\subsection{Volume limited subsamples versus full catalog}
%

According to the above analysis, the choice $\omega=1$ 
approximately minimizes the finite volume error. As there is no
selection effect in an homogeneous catalog, the discreteness error is
minimized as well with $\omega=1$; finally, 
so is the edge effect contribution (\S~3.3). 
This confirms the common wisdom, that
the optimal weight in an homogeneous catalog is $\omega\simeq 1$,
be it volume limited or a two-dimensional galaxy catalog.
%
%

\begin{figure}
\centerline{\hbox{\psfig{figure=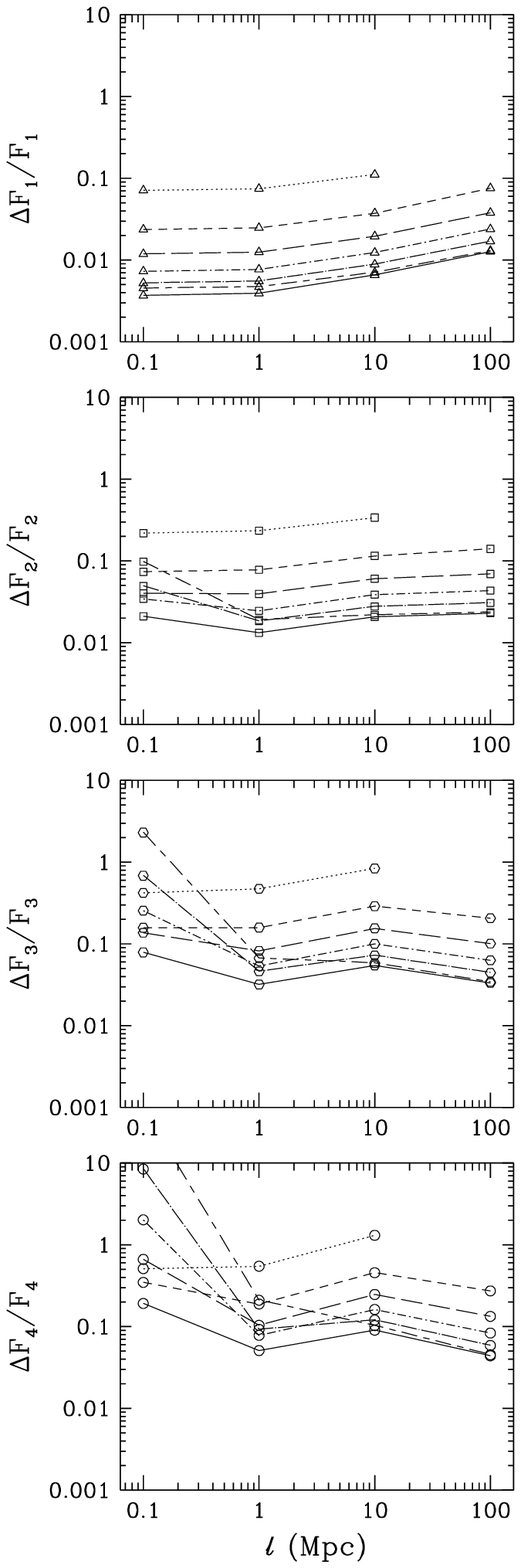,height=19.8cm}}}
\caption[]{The cosmic error is displayed as a function
of scale for our virtual SDSS like catalog ${\cal
E}$ and various volume-limited subsamples ${\cal E}^{i}_{\rm VL}$ (see
\S~4.1), as expected for factorial moments
$F_k$ measured with the optimal weight.
 Each panel corresponds to a value of $k$, increasing from top to bottom.
The solid curves correspond to the parent sample ${\cal E}$. The dots,
short dashes, long dashes, dot-dashes, dots-long dashes, short
dashes-long dashes correspond respectively to the subsamples of depth
$R_i=200$, 400, 600, 800, 1000 and 1200 Mpc. Note, that for $R_i=200$ Mpc, 
there is a point missing for $\ell =100$ Mpc, as a sphere of such
radius is too large to be included in the subsample.}
\label{fig:figure5}
\end{figure}

In figure \ref{fig:figure5}, the cosmic error on the factorial
moments is displayed as a function of scale for our catalog
${\cal E}$ and its volume limited subsamples ${\cal E}^i_{\rm VL}$ defined
in end of \S~4.1. From top to bottom, we have $k=1$, 2, 3 and 4. The
solid lines correspond to ${\cal E}$ with optimal weight.
The dots, dashes, long dashes, dot-dashes, dot-long dashes and long
dashes-short dashes correspond respectively to $R_i=200$, 400, 600, 800,
1000 and 1200 Mpc. 

The figure illustrates clearly that a measurement with
optimal weights using the full catalog 
yields smaller variance than any volume limited subsample
(hereafter VLS). Small VLSs (with small depth) are denser 
than large VLSs (with large depth), so the shot
noise error is more significant on the latter than on the former.
The opposite is true for the finite volume error and the edge effect
error. Large VLS are thus suited
for probing large scales, while small VLS probe
small scales, especially at high order $k$.  
According to this argument, it is possible to construct
a VLS which is fine tuned for a particular scale. At {\em this}
scale, the cosmic error is almost (but not quite) as small as 
the one obtained from the optimal weights from the full survey.
However, this is not true for other scales, i.e. the dynamic range
is quite narrow. Therefore, a series of VLSs has to be constructed,
each optimized for a different scale. As a result, the collection
of VLSs can achieve almost as small errorbars as the optimal
measurement on the full catalog only at the expense of a lot more work.
In summary, a single optimal measurement yields smaller errorbars
more efficiently than a strategy based on a series of VLSs.

%
%
\section{Sparse sampling strategies}
%
%
So far we dealt with the problem of extracting information from existing
catalogs in an optimal way. Another degree of freedom arises, during of
the design of a survey. Next we will be concerned with the optimal
design, especially with the optimal use of the available telescope
time. This in turn inevitably leads to the issues of sparse sampling
and optimal survey geometry. We discuss them as follows:

\begin{enumerate}
\item In the spirit of K86 (see also the recent work
of Heavens \& Taylor 1997), we explore the question: given a fixed
amount of telescope time, how to build a statistically optimal
three-dimensional magnitude limited catalog, if one has the freedom to
sample randomly a fraction $f \leq 1$ of the visible galaxies? As
mentioned in the introduction, a small sampling rate $f$ allows the
construction of a deep but sparse survey. This results in small finite
volume and edge errors, but large discreteness effects. The reverse
is true when $f$ is large. The best compromise between these
requirements yields the optimal sampling rate, which depends on the
scale considered and on the statistic. Here, we extend the calculations of K86
using a more accurate estimate of the cosmic error. While originally
only the two-point function was considered, the optimal sampling rate
will be calculated for higher order factorial moments, $F_k$, $k \leq 4$.
%
%
In \S~5.1, to simplify the analysis, the survey is assumed to have full
sky coverage. The conclusions, however, do not depend significantly on
this assumption. We also suppose that the redshifts are collected
individually. In \S~5.2, the changes brought by multifiber
spectroscopy are discussed, as today this is the most widespread
method for collecting redshifts.
\item  Following K96, \S~5.3 considers the question of optimal survey
geometry. A survey is assumed to cover a fraction of the sky with
redshifts collected by multifiber spectrographs. Thus the catalog can
be naturally decomposed into small patches corresponding to the field
of view of the telescope. The design goal is the optimal arrangement
of these patches. Some of the choices are compact, elongated (or VLA like),
or quasi randomly spread over the sky. This issue is immensely complicated
by several non-linear factors and details of the actual parameters
of the proposed survey. Accordingly, we only attempt to illustrate the
problem, and give an approximate solution under generic
circumstances.
\end{enumerate}

\subsection{Full Sky Survey}
Given the luminosity function of equation (\ref{eq:lumif}), and assuming,
as in K86, that the time required for measuring the redshift of a galaxy 
is inversely proportional to the luminosity
$L$, the total telescope time required 
for constructing a magnitude limited catalog is
\begin{equation}
  T_{\rm total} \propto \int_0^{R_{\rm max}} r^4 dr \phi_* 
	\Gamma[ \alpha, L_{\rm lim}(r)/L_* ],
  \label{eq:totalt}
\end{equation}
where $R_{\rm max}$ is the depth of the survey, and 
the redshifts are assumed to be collected individually.
The magnitude limit $M_{\rm lim}$ of the catalog is related to its
depth $R_{\rm max}$ through 
\begin{equation}
  M_{\rm lim}=5 \log_{10} \left( \frac{R_{\rm max}}{R_{\rm ref}}
  \right) + M_{\rm ref},
\end{equation}
where $R_{\rm ref}$ and $M_{\rm ref}$ are constants. This implies that
\begin{equation}
  T_{\rm total} \propto \phi_* R_{\rm max}^5,
\end{equation}
in agreement with K86. Thus sampling a fraction $f$ of the galaxies
for a fixed telescope time results in 
\begin{equation}
 	R_{\rm max} = R_{\rm ref} f^{-1/5}. 
   \label{eq:rmax}
\end{equation}
Given a scale $\ell$ and a value of the order $k$, the optimal
sampling rate $f$ by definition minimizes the cosmic error of the
factorial moment $F_k(\ell)$. Before any analytical estimates of the
optimal sampling rate, let us consider an example: a full-sky survey
${\cal S}(f)$, with luminosity function and statistics identical to
\S~4.1. We choose
\begin{equation}
   M_{\rm ref}=15.5, \quad R_{\rm ref}=391\ {\rm Mpc}.
   \label{eq:refval}
\end{equation}
For $f=1$ this is roughly a full sky CfA2 catalog, although actually denser 
(see, e.g.~de Lapparent, Geller \& Huchra 1989). On average it contains 
\begin{equation}
   N_{\rm obj}(f=1)\equiv N_{\rm ref} \simeq 72400 
	\label{eq:nobjref}
\end{equation}
galaxies. 

For this hypothetical survey, the optimal sampling rate was found
numerically by calculating the cosmic error as explained in
detail in Appendix~B.3. Figure~\ref{fig:figure6} shows the results for
$F_k$, $1 \leq k \leq 4$, as a function of scale. The symbols
(respectively triangles, squares, pentagons and hexagons for $k=1$,
$2$, $3$ and $4$) take into account effects of the selection function
by using the approximation (\ref{eq:goodapp}) for the optimal weight
$\omega$. The curves (respectively dots, short dashes, long dashes,
dot-dashes for  $k=1$, $2$, $3$ and $4$) have an uniform selection
function: they correspond to a homogeneous sample of same size
$R_{\rm max}(f)$ and involving the same number of objects $N_{\rm
obj}(f)$ as ${\cal S}(f)$. Apart from the small shift for $k=1$, the
curves superpose quite well to the symbols, showing that such an
approximation is valid. This reduces significantly the complexity of
the calculation of the optimal sampling rate. Note, however, that our
choice of $R_{\rm ref}$ [eq.~(\ref{eq:refval})] was not arbitrary:
$R_{\rm ref}$ was chosen to be twice the radius of the volume-limited 
subsample which contains the largest number of objects (for
$f=1$). The following scaling is thus true 
\begin{equation}
    R_{\rm ref}  10^{-0.2 M_{\rm ref}}={\rm constant}\equiv C_{\rm ref}.
\end{equation}
With this choice of $R_{\rm ref}$, the sphere of radius $R_{\rm
max}(f)$ includes most of the detectable galaxies. With a larger
(smaller) value of $R_{\rm ref}$ than given by equation
(\ref{eq:refval}), the curves on figure~\ref{fig:figure6} would be
shifted upwards (downwards). 
%
%
%
%
\begin{figure}
\centerline{\hbox{\psfig{figure=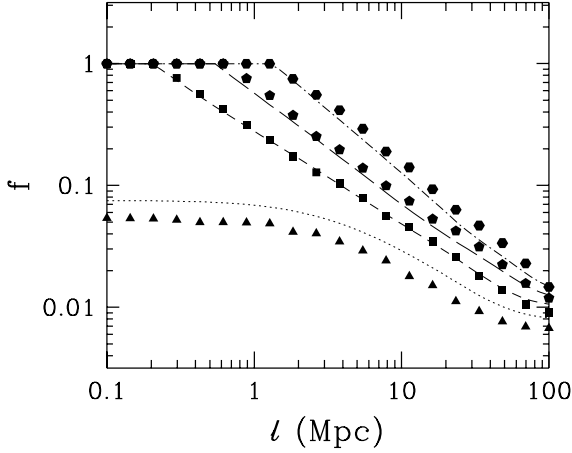,width=8cm}}}
\caption[]{The optimal sampling rate $f$ for measuring the factorial
moment of order $k$ is shown as a function of scale for our
virtual survey ${\cal S}(f)$ (see text). The symbols use a realistic
selection function, whereas the curves assume uniform selection.  
The triangles (dots),
squares (short dashes), pentagons (long dashes) and hexagons
(dot-dashes) correspond to $k=1$, $2$, $3$, and $4$,
respectively.}
\label{fig:figure6}
\end{figure}

\begin{figure}
\centerline{\hbox{\psfig{figure=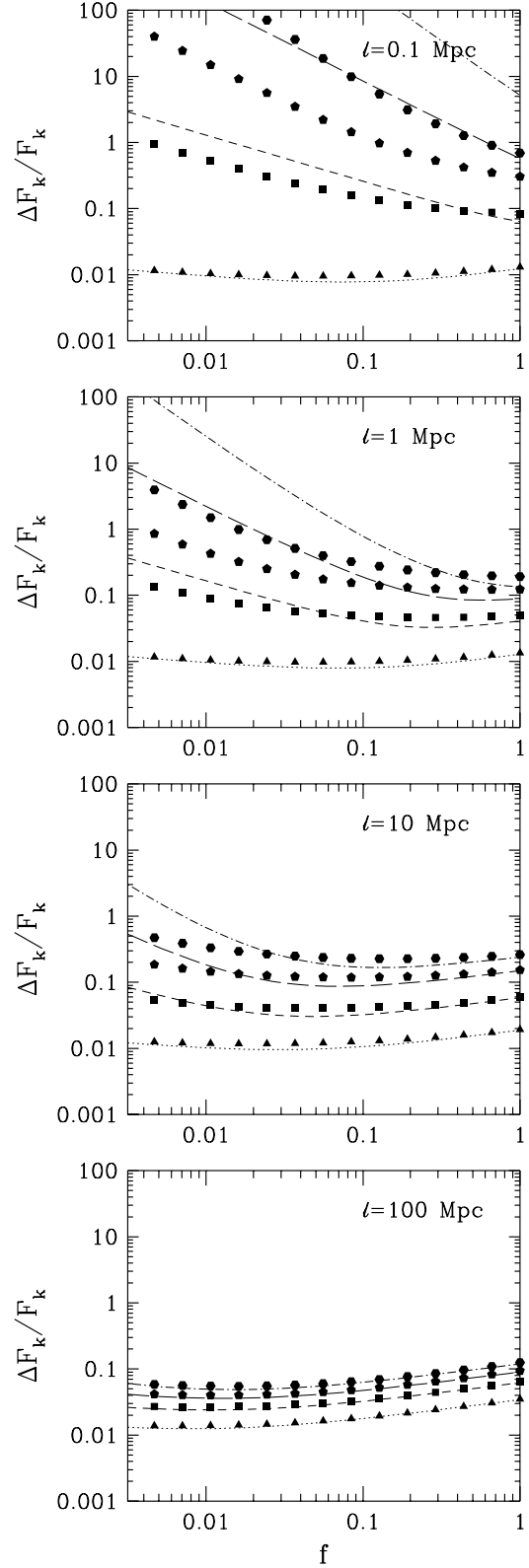,height=21.5cm}}}
\caption[]{The cosmic error on the measured factorial moment of order
$k$ is displayed, 
as a function of the sampling rate $f$ for our virtual survey
${\cal S}(f)$. Each panel corresponds to a given choice of scale
$\ell=0.1$, $1$, $10$ and $100$ Mpc from top to bottom. As in
figure~\ref{fig:figure6}, the symbols use a realistic selection 
function whereas the curves correspond to uniform selection.}
\label{fig:figure7}
\end{figure}

Figure~\ref{fig:figure7} shows the cosmic error on the measured
factorial moments as a function of the sampling rate $f$ for various
scales $\ell=0.1$, $1$, $10$ and $100$ Mpc. The symbols are the same
as in figure~\ref{fig:figure6}. Although it is an excellent
approximation for determining the optimal sampling rate, 
a uniform selection is inaccurate for
estimating  the cosmic error in general, except to some extent
for $k \leq 2$, and for large scales otherwise. 

From figure~\ref{fig:figure6}, the optimal sampling rate
$f$ exhibits a remarkable
power-law behavior up to the saturation to unity, except for $k=1$, when
it rapidly converges to a value smaller than unity at small scales. Moreover, 
$f$ increases with the order $k$, corresponding to
the increasing relative contribution of the discreteness error
with $k$. These features can
be further explained by analytical calculations as follows.

According to the previous findings, calculations will be simplified
by assuming a uniform selection. This appears to be good approximation
for determining $f$. The analytical formulae of SC yield the relative
cosmic error on the measurement of $F_k$, $k \leq 3$. This can be
used to find the optimal sampling rate in the
weakly and highly non-linear regimes. The
details of the calculations can be found in appendix~C. 

In the highly nonlinear regime, ${\bar \xi} \gg 1$, i.e.~at small
scales, edge effects are expected to be negligible compared to finite
volume effects (e.g.~SC). The optimal sampling rate yields the best
compromise between discreteness and finite volume
effects. If $R_{\rm ref}$ is sufficiently large, ${\bar \xi}(R_{\rm
max})$ is expected to exhibit an approximate power-law behavior. For
a Harrison-Zeldovich power-spectrum $\langle |\delta_k|^2 \rangle
\propto k$, 
\begin{equation}
  \xiav(R_{\rm max}) \simeq \left( R_{\rm max}/L_0
  \right)^{-\gamma_L}, \quad \gamma_L=4.
  \label{eq:xibo}
\end{equation}
The power-law behavior of the average correlation function
in the nonlinear regime is a good approximation [eq.~(\ref{eq:xipow})],
and, although not absolutely necessary,  simplifies the computations:
\begin{equation}
 {\bar \xi}(\ell) =
 (\ell/\ell_0)^{-\gamma}, \quad \ell \la \ell_0.
\end{equation}
With the above hypotheses, the optimal weight, when not saturated to
unity, is
\begin{eqnarray}
  f & \simeq & \displaystyle \left[ \frac{5 k-3}{\gamma_L} 
  \frac{\alpha_k^{\rm D}}{\alpha_k^{\rm F}} 
  \frac{N_{\rm ref}^{-k} R_{\rm ref}^{\gamma_L+3(k-1)}}
  {L_0^{\gamma_L}\ell_0^{\gamma(k-1)}}
  \right]^{\frac{5}{5k
  +\gamma_L-3}} \ell^{-\frac{5(3-\gamma)(k-1)}{5k+\gamma_L-3}},
  \nonumber \\
  & &{\bar \xi} \gg 1. 
  \label{eq:foptima}
\end{eqnarray}
In this expression, the quantities $\alpha_k^{\rm D}$ and
$\alpha_k^{\rm F}$ are numbers depending on the order $k$ and on
$Q_l$, $l \leq 2k$. Their ratio writes, for $k \leq 3$,
\begin{equation}
   \frac{\alpha_1^{\rm D}}{\alpha_1^{\rm F}}=1,
   \quad \frac{\alpha_2^{\rm D}}{\alpha_2^{\rm F}}\simeq \frac{1}{8
   Q_4},
   \quad \frac{\alpha_3^{\rm D}}{\alpha_3^{\rm F}}\simeq
   \frac{Q_3}{47.4 Q_6}.
\end{equation}
Equation~(\ref{eq:foptima}) is not valid for the Gaussian case, except
for $k=1$. Then $f$ is independent of scale when ${\bar \xi} \gg 1$,
explaining the result obtained for the triangles and the dots in 
figure~\ref{fig:figure6}.

In the weakly nonlinear regime, ${\bar \xi} \ll 1$, i.e.~at large
scales, edge effects typically dominate over finite volume
effects (e.g.~SC). Thus the optimal sampling rate $f$ results from a
competition between edge effects and discreteness effects (see
Appendix~C for the details). At large scales, ${\bar \xi}$ is not a
power-law, (except for {\em very} large scales, $\ell \ga 100$ Mpc),
neither is the optimal sampling rate  
\begin{equation}
  f=\frac{2 k^2}{3 \alpha_k^{\rm E}} \frac{R_{\rm ref}^3}{N_{\rm ref}}
  \frac{1}{{\bar \xi} {\ell^3}},\quad {\bar \xi} \ll 1,
	\label{eq:foptima2}
\end{equation}
where $k^2/\alpha_k^{\rm E}$ is slowly
increasing with  $k$:
\begin{equation}
   \frac{1}{\alpha_1^{\rm E}}\simeq 0.18,\quad \frac{4}{\alpha_2^{\rm
   E}} \simeq 0.23, \quad  \frac{9}{\alpha_3^{\rm E}}\simeq 0.26.
\end{equation}

For $\gamma < 3$ the optimal weight is a decreasing function of
scale. It depends on the order $k$ considered, and, for nonlinear
scales, on the details of the higher order statistics through the
ratio $\alpha_k^{\rm D}/\alpha_k^{\rm F}$. However, it has a weak
dependence on the total telescope time $T_{\rm total}$ according to
the following argument: $T_{\rm total} \propto R_{\rm ref}^5$, $N_{\rm
ref}\propto R_{\rm ref}^3$, therefore in the highly nonlinear regime
[eq.~(\ref{eq:foptima})], 
\begin{equation}
   f \propto T_{\rm total}^{\frac{\gamma_L-3}{5 k + \gamma_L-3}},
  \quad {\bar \xi} \gg 1,
\end{equation}
constituting a weak dependence of the optimal sampling rate
on the total observing time, at least for $k \geq 2$. In the weakly
nonlinear regime, ${\bar \xi} \ll 1$, $f$ does not depend at all on
$T_{\rm total}$. Consequently, the results displayed in
figure~\ref{fig:figure6} are more general than initially suspected: they
should be roughly valid for any full sky survey of
depth larger than a few hundred Mpc (so that ${\bar \xi}(R_{\rm max})$ is
approximately a power-law), to the extent that the assumptions we have
made for the underlying statistical properties (see \S~4.1) are realistic. 

To give orders of magnitude, let us compute numerically what would be
typically the optimal sampling rate from formulae (\ref{eq:foptima})
and (\ref{eq:foptima2}). Assuming $H_0=50$ km/s/Mpc, a reasonable
choice for the correlation length in equation (\ref{eq:xibo}) is $L_0
\simeq 50$ Mpc. The same values are taken as in equations
(\ref{eq:refval}) and (\ref{eq:nobjref}) for $R_{\rm ref}$ and $N_{\rm
ref}$. With $\gamma=1.8$, $\ell_0=16$ Mpc and the values of $Q_N$ quoted 
in equation~(\ref{eq:qns}), from equation (\ref{eq:foptima})
the optimal  weight $f_k$ is obtained 
corresponding to each value of 
$k$ in the highly nonlinear regime:
\begin{eqnarray}
   f_1 & \sim & 0.05, \\
   f_2 & \sim & \min(1, 0.2 \ell^{-0.55}), \label{eq:f22}\\
   f_3 & \sim & \min(1, 0.3 \ell^{-0.75}). 
\end{eqnarray}
Setting $\ell=10$ Mpc in equation (\ref{eq:f22}) gives $f_2 \sim
1/18$, in rough agreement with figure~\ref{fig:figure6}.
This result is similar to the findings of K86. This is not surprising,
although his calculation was done for rather larger scales $\ell \sim
30$ Mpc. In the weakly nonlinear regime, equation (\ref{eq:foptima2}) 
must be used. For a CDM spectrum normalized to COBE
(${\bar \xi}(16\ {\rm Mpc}) \simeq 1.22^2$, see, e.g.~Bunn \& White (1997))
and $\ell=100$ Mpc, $f_2 \sim 1/135$ would be
optimal: a quite small sampling rate, in rough agreement with
figure~\ref{fig:figure6}. 

More importantly, however, the sampling rate has to be optimized
for range of scales and statistics, not only for fixed values of $k$
and $\ell$. The optimal sampling rate for $(k,\ell)=(2,100)$, $f \sim
1/130$, is not optimal for other scales and orders: according to
figure~\ref{fig:figure7} (symbols), it dramatically penalizes
small scales, especially when $k$ is large: for $k=4$ and $\ell=1$
Mpc, the error is about one order of magnitude larger than for $f=1$.
Conversely, $f \simeq 1$ is a good choice for small
scales, however it would increase the errors on large scales
by a factor of $2\div 3$. A good compromise seems to be $f$ of order
$1/10$, in approximate agreement with the initial findings $f=1/20$ of
K86. This choice unfortunately only optimizes for low-order
statistics, $k \leq 4$. The higher is $k$, the higher should be
$f$. In particular, if one wants to analyze the probability distribution
function of the density field in terms of shape (e.g.~Bouchet et
al.~1993, CBSII), perform cluster selection (Szapudi \& Szalay 1993a, Szapudi
\& Szalay 1996), or any other investigation which depends on the full
hierarchy of the factorial moments, a sparse sampling strategy seems
inappropriate (as already stated in CBSII). In summary, although for
low order moments a sampling rate $f \leq 1$ can be found, which
globally optimizes the measurement, higher order moments require full
sampling. This conclusion is further supported by the arguments of
the next section, where it is shown that, with the advent of multifiber
spectroscopy, there is more to be lost by undersampling than any
possible gain.  

Previously, we assumed that the telescope time $t_{\rm obs}$
required to measure the
redshift of a galaxy of luminosity $L$ was proportional to $1/L$. This
assumption probably breaks down for faint objects when the 
background noise dominates. Then the scaling
$t_{\rm obs} \propto 1/L^2$, as used by Heavens \& Taylor (1997),  
would probably be more appropriate. Using this scaling does
not significantly change the previous results and conclusions.
The greatest difference is at large scales, 
where the optimal sampling rate becomes at most
twice as small compared to figure~\ref{fig:figure6},
where $t_{\rm obs} \propto 1/L$ was 
assumed\footnote{This comparison supposes that the optimal sampling rate
does not depend on total telescope time.}. 
Note also that when $t_{\rm obs} \propto 1/L^2$, assuming
uniform selection is not seen to be a good approximation anymore, thus
analytical estimates similar to equations (\ref{eq:foptima}) and
(\ref{eq:foptima2}) are quite inaccurate.
%
\subsection{Multifiber spectroscopy}
%
The above calculation was assuming that the redshifts are collected
individually. In today's astronomy, the problem is complicated by the
fact that multifiber spectrographs are used to collect redshifts. With
this technique it is possible to measure simultaneously $N_{\rm S}$
redshifts in a patch of the sky of fixed angular size $\Omega_{\rm
S}$. This limits the validity of the previous calculations to the case
when the typical number of candidates in such a field, $N_{\rm
can}\propto f^{2/5}$, is larger than $N_{\rm S}$, as already discussed
in K86\footnote{By taking into account the galaxy density
fluctuations from patch to patch over the sky, a more realistic
constraint would be $N_{\rm can}[1-\sqrt{1/N_{\rm can}+w(\Omega_{\rm
S})}] \ga N_{\rm S}$, where $w(\Omega_{\rm S})$ is the average of the
angular correlation function over the patch. This condition accounts
for the expected Poisson fluctuations and intrinsic correlations of
the density field, up to second order.}.

Even if the available telescope time is too short to allow $N_{\rm
can} \ga N_{\rm S}$, a sparse sampling strategy still can make sense.
It is possible to spend more time per field in order to go deeper, and
thus allow a smaller value of $f$. There is a price to pay for such
strategy: if $T_{\rm total}$ is fixed, the total solid angle $\Omega$
covered by the survey will be reduced. More details of this argument
can be found in K86. Here let us only remind the scaling of the
relevant quantities with the sampling rate, such as the depth $R_{\rm
max}$ of the catalog, its solid angle $\Omega$ and its volume $V=\Omega
R_{\rm max}^3/3$: 
\begin{equation}
  R_{\rm max}=R_{\rm ref} f^{-1/3},\quad \Omega=\Omega_{\rm ref}
f^{2/3}, \quad V=V_{\rm ref} f^{-1/3}.
\end{equation}
Similarly to K86, we are going to see how these new relations
affect the optimal sampling rate, compared to \S~5.1. 
Just as before, uniform selection will be assumed, because it was
shown that the details of the selection function do not
influence the results. Unfortunately, the varying shape of the catalog
renders the calculation of the finite volume error costly: for each
value of $f$, $\Omega$ changes, and a new calculation of the two-point
correlation matrix ${\tilde \xi}(r_1,r_2)$ is required
[eq.~(\ref{eq:defxit})]. This would go beyond the scope of this
paper. Instead the finite volume error is simply calculated for a
spherical catalog of same volume $V$. This is a reasonable approximation
if the catalog is compact enough, although it would obviously fail
for the extreme case of a pencil beam survey. We shall partly come
back to that problem in \S~5.3.

Similarly to \S~5.2, and after the calculations detailed in Appendix
C, the result in the highly nonlinear regime is 
\begin{eqnarray}
  f & \simeq & \left[ \frac{9 k-3}{\gamma_L} 
  \frac{\alpha_k^{\rm D}}{\alpha_k^{\rm F}} 
  \frac{N_{\rm ref}^{-k} L_{\rm ref}^{\gamma_L+3(k-1)}}{L_0^{\gamma_L}\ell_0^{\gamma(k-1)}}
  \right]^{\frac{9}{9k
  +\gamma_L-3}} \ell^{-\frac{9(3-\gamma)(k-1)}{9k+\gamma_L-3}},
  \nonumber \\
  & & {\bar \xi} \gg 1,
  \label{eq:foptimabis}
\end{eqnarray}
where the typical size $L_{\rm ref}$ of the
catalog is defined as
\begin{equation}
   V_{\rm ref} \equiv \frac{4\pi}{3} L_{\rm ref}^3.
\end{equation}
In the weakly nonlinear regime, ${\bar \xi} \ll 1$,
the edge effects are likely to dominate over finite volume effects.
To leading order in $v/V$ the solution is insensitive
to the shape of the catalog, although as shown in \S~5.3, 
this is not true in extreme cases. The optimal sampling rate is
\begin{equation}
  f=\frac{2 k^2}{\alpha_k^{\rm E}} \frac{L_{\rm ref}^3}{N_{\rm ref}}
  \frac{1}{{\bar \xi} {\ell^3}}, \quad {\bar \xi} \ll 1.
	\label{eq:foptima2bis}
\end{equation}
Equations (\ref{eq:foptimabis}) and (\ref{eq:foptima2bis}) are quite
similar to eqs.~(\ref{eq:foptima}) and (\ref{eq:foptima2}) in
\S~5.1. For example, in the weakly nonlinear regime, the optimal
sampling rate is roughly three times larger for multifiber
spectroscopy, than in \S~5.1 (instead of the factor
two of K86). In the highly nonlinear regime, the difference is 
apparently smaller, although this is probably partly due to our
way of estimating the finite volume error, and changes slowly with
scale. For multifiber spectroscopy, $f$ is typically twice than for
individual collection of redshifts for $k=1$, $30\%$ more for $k=2$
and $20\%$ more for $k=3$. In the weakly nonlinear regime, our 
estimate of the cosmic error is accurate enough to compute the gain
on the errors for optimal compared to full sampling, even though 
uniform selection was used (see lower-panel of Fig.~\ref{fig:figure7}).
For a CDM spectrum normalized to COBE at $\ell=100$ Mpc with the
corresponding optimal sampling rate $f_2 \sim 1/45$, the gain is
merely $1.5$ in reduction of $\Delta F_2/F_2$. This is to be compared
with $2.8$ for individual redshifts. These results qualitatively agree
with K86: the optimal sampling rate is increased with multiple
collection of redshifts compared to individual collection, and the
corresponding gain in the cosmic error is smaller. This strengthens
the conclusions of \S~5.1: even if a sparse sampling strategy could be
optimal for special applications, such as measuring low-order
statistics, the resulting gain is too small compared to the
corresponding increase of errors for high order statistics which are
more sensitive to sampling. Another important point to note is that
sparse sampling strategy consists of measuring deeper redshifts, thus
less controllable systematic errors are also likely to increase,
which are not included in the previous discussions.
%
\subsection{Survey geometry}
%
So far we neglected the dependence of the cosmic error and the
corresponding optimal sampling rate $f$ on catalog geometry: this
amounts to replacing the geometry with a sphere of same volume. For a
deep catalog covering only a small part of the sky, such as a pencil
beam survey, this is clearly not a good approximation. The finite
volume error depends significantly on the catalog geometry. So does
the shot noise error\footnote{through ``edge-discreteness'' effects as
mentioned in the introduction.}, and particularly the edge effect
error, especially when the cell size becomes comparable to the size of
the largest cell contained in the catalog. This did not show up in our
calculations to leading order in $v/V$: in that case, the edge effect
and the shot noise errors do not depend on the catalog geometry.

Since geometry is another degree of freedom in the design of surveys,
it can be used to achieve predetermined statistical goals. For typical
redshift surveys, the geometry has an ``atomic'' building block,
$\Omega_{\rm S}$, corresponding to the field of view of the
telescope. Within this, a certain number of spectra can be
collected. The design problem which will be discussed in this section
is (K96): what is the optimal distribution of fields of size
$\Omega_{\rm S}$ on the sky? Note that this problem is relevant even
to redshift surveys, which, as the SDSS, plan to uniformly cover a
large portion of the sky. Since such a survey takes a long time to
carry out, it is important that in the initial phases the individual
fields should be placed in such a manner that as much preliminary
information as possible could be extracted. Such strategy can not only
provide preliminary results before the full completion of the survey,
but it gives an excellent early check, whether the envisioned goals of
the survey can be achieved when completed. Note that in such a case,
the design problem acquires another dimension, i.e.~time, since
analysis can be performed at several stages before full
completion. Generalization of the following arguments for this is trivial.
Some of the possibilities for the geometry are a compact structure, an
elongated one (perhaps VLA-like), or a (quasi) random distribution,
etc. Judging all possibilities is extremely difficult, not to mention
that for realistic surveys other ``worldly'' factors play important
roles, such as weather, dark time, season, etc., therefore the aim is
here only to illustrate the problem and present some practical
suggestions by solving it under simplified circumstances. 

Each field corresponds to an elementary galaxy catalog
of volume $V_{\rm S}=\Omega_{\rm S}R_{\rm max}^3/3$ at position ${\hat
n}_i$ in the sky. The centrally important quantity for
the finite volume error, the integral
of the two-point function over the whole survey can be decomposed as
\begin{equation}
  {\bar \xi}(L) = \frac{1}{M_{\rm S}^2} \sum_{i \neq j} w(\theta_{ij})
  +\frac{1}{M_{\rm S}} {\bar \xi}(L_{\rm S}).
  \label{eq:finitepatch}
\end{equation}
In this equation, $M_{\rm S}$ is the total number of fields.
The angular correlation function $w(\theta)$ is defined by
\begin{equation}
   w(\theta)\equiv \frac{1}{V_{\rm S}^2}\int d^3r_1 \int d^3r_2
   \xi(r_{12}),
\end{equation}
where each integral is performed on disjoint elementary catalogs
with positions ${\hat n}_1$ and ${\hat n}_2$ on the sky such that
$\theta=({\hat n}_1,{\hat n}_2)$. The quantity ${\bar \xi}(L_{\rm S})$
is the average of $\xi$ over an elementary catalog: 
\begin{equation}
  {\bar \xi}(L_{\rm S}) \equiv \frac{1}{V_{\rm S}} \int_{V_{\rm S}}
  d^3r_1 d^3r_2 \xi(r_{12}).
\end{equation}
According to equation (\ref{eq:finitepatch}), values of $\theta_{ij}$
could be chosen, which minimize ${\bar \xi}(L)$ and therefore the
finite volume error. This choice would not affect the
edge effect error and the shot noise error significantly: they do not
depend on the geometry of the catalog to leading order in $v/V$,
i.e.~at small scales.  Although the solution depends on the large
scale behavior of the two-point function, we conjecture the best
configuration to be a ``glass'' spread over the largest possible area
of the sky. This configuration would maximize the distance between the
fields, and, in agreement with intuition, it would decrease the
coherence of the fields as much as possible. Clearly, the ``compact''
configuration is the worst, followed by the ``line''
configuration. These conclusions are very similar to K96, who
discussed optimal survey strategies for measuring weak lensing related
two-point statistics in Fourier space.  

However, the above arguments are relevant only if the considered scale
$\ell$ is small compared to the size $\ell_{\rm max}$ of the largest
cell included in the catalog. When $\ell$ becomes comparable to
$\ell_{\rm max}$, our calculations of the full cosmic error are not
valid anymore, although the conclusion remain approximately valid for
the finite volume error. The exact calculation of the cosmic error in
the regime where $\ell \la \ell_{\rm max}$ is quite tedious (see
appendix B of SC for an example). Qualitative description, however,
can be given. The shot noise error and the edge effect error increase
with $v/{\hat V}$, where ${\hat V}$ is the volume occupied by
positions of cells {\em included} in the catalog. This ratio increases
toward larger scales both because $v$ is increasing and ${\hat V}$
decreasing. This latter effect is obviously more prominent for
``glass'' , or ``line'' configurations. The edge effect error (and the
shot noise error) is therefore more important at large scales for
these configurations, than for the compact geometry.

As before, the choice of the optimal geometry of the catalog
results from the competition between various contributions to the
cosmic error, namely the finite volume error against the edge
effect and the shot noise errors. The details of the sampling strategy
depend on the value of $\Omega_{\rm S}$, on total telescope time, the
needed dynamic range in scales, the statistical aim,  and the
clustering properties of the universe. A good compromise could be a
survey composed of several compact subsamples
of size $\Omega_{\rm F} \geq \Omega_{\rm S}$, 
spread over a large fraction of the sky  on a
glass like structure. The optimal value of $\Omega_{\rm F}$ is a
complicated function of the various parameters of the survey. While
the details of this difficult issue are left for subsequent research,
it is clear that the above arguments give an approximate solution once
the survey parameters and the scale range are fixed.
%
%
\section{Discussion}
%
%
In this article, we examined extensively the measurement of low-order
moments of the count probability distribution function (CPDF) in
three-dimensional magnitude limited galaxy catalogs, with special
emphasis on issues related to the effects of the selection function.
A new estimator was proposed: the weighted factorial moment of
count-in-cells, corrected for selection effects
[eq.~(\ref{eq:myw})]. The following questions were studied in detail:
\begin{enumerate}
\item[(i)] Given a catalog, what is the optimal way of measuring
factorial moments?
\item[(ii)] What is the optimal sampling
strategy for constructing a catalog to measure factorial moments?
\end{enumerate}
Both of these question are intimately related to the variance of the
proposed unbiased estimator. Thus the cosmic error was computed,
by  extending the calculations of Szapudi \& Colombi (1996, SC) 
for the new estimator which
includes a local statistical weight, and for a general selection
function. Similarly to SC, the hierarchical model and local Poisson
behavior was used to simplify the calculations. The local (but not
the global) variations of the selection function were neglected
as well. This allowed for the first time the accurate estimation
of the optimal weight, i.e.~the one which minimizes the cosmic error
[eqs.~(\ref{eq:cosmic2}) to (\ref{eq:optiw2})].

To illustrate numerically the first question (i), a virtual SDSS-like
catalog was considered. For this, we demonstrated the advantages of
our new estimator, which extracts all the relevant information from
the catalog at once. This method not only yields higher accuracy in a
wider dynamic range than the more traditional volume limited method,
but it is significantly more efficient as well. As an added benefit,
our calculation of the cosmic error finds the best volume limited
strategy if other reasons necessitate its use. A remarkably simple
expression for the optimal weight $\omega$ was found, which provides
an excellent approximation to the solution of the corresponding
integral equation
\begin{equation}
	\omega(r) \propto 1\left/ \left[ \Delta^2_{\rm F}
	+\Delta^2_{\rm E}+\Delta^2_{\rm D}(r) \right] \right.,
	\label{eq:goodapp2}
\end{equation}
where $\Delta^2_{\rm F}$, $\Delta^2_{\rm E}$ and $\Delta^2_{\rm D}(r)$
are respectively the finite volume, the edge effect and the
discreteness errors. As a result, the optimal weight for an
homogeneous sample, i.e. with uniform selection function, is very well
approximated by $\omega=1$. Unlike $N$-point correlation functions, it
is impossible to correct for edge effects with our estimator, due to
the finite extension of the smoothing kernel. A different approach,
however, is presented elsewhere (Szapudi \& Szalay 1997b).

Interestingly, it appears that $\omega=1$ is a good approximation for
the optimal weight in the SDSS catalog, at scales larger than $\sim 1$
Mpc. As illustrated by figure~\ref{fig:figure4}, the corresponding
cosmic error on the estimates of $\langle \rho^k \rangle$ should be
rather small, of order $1-2\%$ for $k=2$, $3-5\%$ for $k=3$ and
$5-10\%$ for $k=4$. These results, however, depend on the details of
the model we used for clustering properties of the universe.

Note the similarity of our results with those of Feldman, Kaiser \&
Peacock (1994, hereafter FKP), who did similar calculations for
finding the optimal weight to measure the power-spectrum $P(k) \equiv
\langle |\delta_{k}|^2 \rangle$, where $\delta_{k}$ is the Fourier
transform of the density contrast $\delta=\rho-1$. They assume that
the modes are normally distributed in Fourier space and consider modes
corresponding to scales small compared to the size of the
catalog. Their result is  
\begin{equation}
 \omega(\r) \propto \frac{1}{1/n(\r) + P(k)},
\label{eq:optiwk}
\end{equation}
in excellent agreement with equation (\ref{eq:goodapp2}). Indeed, from
equation (\ref{eq:optiwk}), the optimal weight presents a plateau at
small $r$ and decreases exponentially at large $r$. The same
conclusions apply as well if we compare our optimal weight to the 
one obtained for measuring optimally the two-point correlation
function $\xi(r)$ (e.g.~Efstathiou 1996; Hamilton 1997a).

To address question (ii), we considered the sparse sampling strategy
proposed by Kaiser (1986, K86, see also Heavens \& Taylor 1997). 
This consist of randomly sampling a
fraction $f$ of the visible candidates to build a three dimensional
catalog. The optimal sampling rate $f$ minimizes the cosmic error,
when the total available telescope time is fixed. A small sampling
rate $f$ results in a sparse but deep catalog, therefore decreasing
the finite volume and the edge effects. Then discreteness error
becomes the limiting factor. The reverse is true for
high sampling rate. To measure factorial
moments up to order four in a reasonable range of scales such as $1$
Mpc $\la \ell \la$ 100 Mpc, $f=1/10-1/3$ yields a good compromise
between the above effects, depending whether the redshifts are
collected individually, or, as usual today, collectively. This is in
qualitative agreement with K86, although slightly larger for the
following reasons: our calculation of the cosmic error is more
realistic; we consider higher order (4 vs.~2) statistics than K86; we
emphasize dynamic range from small scales to large scales, the latter
being solely considered by K86. Note that the optimal sampling rate
increases with the order $k$ of the statistic considered and decreases
with scale\footnote{Except for very large scale $\ell \ga$ a few
hundred Mpc. In this regime, one expects ${\bar \xi} \propto \ell^{-4}$ 
[eq.~(\ref{eq:xibo})] implying $f \propto \ell$
[eqs.~(\ref{eq:foptima2}), (\ref{eq:foptima2bis})], 
in agreement with the calculations of Heavens \& Taylor (1997)
concerning the power-spectrum.}.
Therefore, when properties sensitive to higher than fourth
order are considered, such as the CPDF shape in the nonlinear regime, 
cluster selection, higher,
most likely full sampling will be optimal (Colombi et al.~1995,
Szapudi \& Szalay 1996), all the more since we have shown that a
sparse sampling strategy yields only a marginal gain compared to full
sampling even at the largest scales. 

Finally, the design of optimal geometry for a galaxy survey covering
some fraction of the sky (e.g.~Kaiser 1996, K96) was considered. This
is an important problem even for surveys eventually covering a large,
continuous portion of the sky, since early results can be obtained by
the appropriate (evolving) geometry. Again, the design is governed by
the competition between edge and discreteness effects, requiring
compactness at large scales, and finite volume effects, requiring
large sky coverage. For a reasonably deep survey, where the finite
volume error is expected to be quite small, the compact geometry is
probably the best choice, because it allows the largest possible scale
range for the measurement of the moments. A glass like configuration,
where the survey is spread out over the sky in pencils beam
subsamples, would reduce the finite volume error. It would be,
however, suboptimal for edge effects at large scales, thus
constraining the dynamic range of the measurement. A possible
compromise is to increase the size of each subsample, for compactness,
and still spread them over the sky in a quasi random glass like structure,
for reducing the finite volume effects. The details of such a
construction, however, depend so much on the goals of the survey in
terms of scale range, desired statistics, and predetermined conditions
as well, such as the field of view of the telescope, the number of
fibers of the multifiber spectrograph, and finally on the statistics 
of clustering in the universe, that we only attempted to
illustrate the problem by solving it for the conceptually simplest
cases.

While the calculation presented in this work provides sufficient
details for most practical applications, there are several points
where generalization or extension could provide more accuracy, if
needed:
\begin{enumerate}
\item 
We did not investigate the dependence of the results on the details of
the clustering in the universe, since a particular model was assumed
to illustrate our method (see \S~4.1). As shown by SC, the cosmic
error on statistics of order $k$ depends on statistics of order
$l \leq 2k$. 
The dependence becomes stronger at small scales,
possibly altering the optimal weight $\omega$. At very large scales, where
the non-Gaussianity is less important, the results are only weakly
sensitive to higher order statistics. At the level of approximation
used in this paper, however, our qualitative results both on optimal
weight and sampling rate are expected to be valid for a broad spectrum
of realistic statistics. Extended perturbation theory (Colombi et
al.~1997) can be used to estimate the sensitivity of the results
within reasonable limits. An overall amplification for the optimal
sampling rate of a factor 4 is predicted for the variance at
nonlinear scales when one passes from an effective spectral index
$n_{\rm eff}=-9$ to $n_{\rm eff}=-1$ (corresponding respectively to
scale-free initial conditions $\langle |\delta_k^2| \rangle \propto
k^n_{\rm linear}$ with $n_{\rm linear}=-2$ and $n_{\rm
linear}=0$). The calculations in this paper correspond to an
intermediate value $n_{\rm eff} \sim -3$ (corresponding to $n_{\rm
linear}=-1$), which is most supported by observations.  
\item So far, we employed the hierarchical tree model as an approximation
to simplify the calculations of the cosmic error. Another possibility is to
use tree-level perturbation theory predictions (see, e.g.~Bernardeau
1994, 1996), which are seen from $N$-body simulations to be valid in
the regime ${\bar \xi} \la 1$ (e.g.~Juszkiewicz,
Bouchet \& Colombi 1993; Bernardeau 1994; Baugh et al.~1995; 
Gazta\~naga \& Baugh 1995; Colombi et al.~1996, 1997).
The use of perturbation theory
predictions for computing the cosmic error will be presented elsewhere
(Szapudi, Bernardeau \& Colombi, 1997, hereafter SBC). 
\item 
As mentioned in the introduction, redshift distortion was
completely disregarded, although it affects the measured statistics in
two different ways, depending on the scaling regime considered. At large
scales, coherent flows enhance the density contrast along the line of
sight, therefore increasing the amplitude of the $N$-point correlation
functions. At small scales, on the contrary, the ``finger of god''
effect by the high velocity dispersion of large clusters
tends to smear out clustering and reduce the amplitude of the
$N$-point correlation functions. The net effect on the parameters
$Q_N$ is a decrease at small scales and no change at large scales 
(e.g.~Matsubara \& Suto 1994; Hivon et al.~1995). One consequence is
that, at variance with what is expected from real space measurements
in $N$-body simulations, function $Q_N(\ell)$ is fairly flat in
redshift space. As a result, the hierarchical model used along this
paper appears to be an even better approximation for estimating the
cosmic error in redshift surveys. However, a correction for redshift
distortions would be needed to transform the measured moments into
configuration space, which is left for future work. 
\item 
Possible dependence of the clustering on luminosity, and morphology
was neglected as well. If clustering of galaxies increased with their
luminosity (as some observations might indicate), magnitude limited
surveys would have inhomogeneous clustering properties. Indeed, only
bright objects are seen deep in the catalog, thus clustering would
change with depth implying a systematically increasing bias with
distance from the observer. In that case, our estimator
[eq.~(\ref{eq:myw})] would have to be corrected for this bias,
otherwise the volume limited approach is the only alternative. In
fact, if the bias is unknown a priori, the volume limited approach
provides a way to estimate it. Using an already existing
formalism  (e.g.~Bernardeau \& Schaeffer 1992), one can probably
directly include the effects of biasing in the calculation of the
cosmic error, but this is left for future work. A similar effect is
caused by the change of clustering with $z$ in a very deep catalog 
(Suto 1997). 
%
\item It is worth to emphasize again, that our purely statistical
approach cannot account for systematic observational errors,
introduced by the instruments, uneven sky, seeing, emission from our
own galaxy, inappropriate K-corrections, problems with star galaxy
separation, inaccuracy of redshifts, etc. It is only hope, that in a
well controlled sample, these effects do not constitute the dominant
source of error, or, turning it around, this condition yields well
defined accuracy goals for observations. 
\item There are known alternatives over constructing an estimator 
$\tF_k$ from the CPDF. For instance, for many applications, the
connected moments, or cumulants $S_N$ (or, equivalently $Q_N=S_N /
N^{N-2}$) are desirable. These numbers are ratios of combinations of
factorial moments (for example $S_3 = \langle \delta^3 \rangle /
\langle \delta^2 \rangle^2$).  Kim \& Strauss (1997) proposed an
alternate method to measure $S_N$ in galaxy catalogs, by fitting the
Edgeworth expansion (e.g.~Juszkiewicz et al.~1995) convolved with a
Poissonian to the measured CPDF. 
They find that this method, less sensitive to the tails of
the CPDF, is more robust than the traditional moment method which was
refined in this paper. However, their error estimation is quite ad-hoc, 
even if normalized with $N$-body simulations. Moreover, their method
is valid only in the weakly nonlinear regime ${\bar \xi} \la 1$,
and, in its present form,  can be applied only to volume limited
catalogs, in contrast with our indicator $\tF_k$.
It would be interesting to apply the formalism developped 
in this paper directly to the cumulants $S_N$ and see how the results
differ from what we obtained for the factorial moments (SBC).
\item The CPDF depends on the full hierarchy of factorial moments. 
In a finite galaxy catalog, however, only a limited amount of
information is available, due to the cosmic error and other sources of
noise. Thus data compression must be possible without loosing
significant information. For example, information content in the tails
of the measured CPDF is small because of large fluctuations associated
to rare events (e.g.~Colombi et al.~1994). Data compression is
already used for applying maximum likelihood tests to the galaxy
distribution and the cosmic microwave background (e.g.~Bond 1995;
Vogeley \& Szalay 1996; Tegmark, Taylor \& Heavens 1996; Jaffe, Knox
\& Bond 1997). For counts-in-cells statistics, such a compression is
complicated by the non-Gaussanity of the likelihood function at small
and intermediate scales (e.g.~SC).  
\item A problem somewhat related to (vi), 
the estimators $\tF_k$ are clearly not statistically independent
(i.e. $\langle \tF_k \tF_{k'} \rangle \ne F_k F_{k'}$ if $k \neq k'$,
$\langle \tF_k \tF_{k'} \tF_{k''}\rangle \ne F_k F_{k'} F_{k''}$, $k
\neq k' \neq k''$, etc.), therefore they do not provide independent
tests of theoretical models.
As a result, the question of determining
the best higher order statistics estimators beyond the two point correlation 
function (or its Fourier transform, the power spectrum) remains opened.
Ideally, one would like to create hierarchies of statistically
independent estimators. Technics to build such hierarchies
are already extensively developped 
for estimating two-point statistics in the galaxy 
distribution and in the cosmic microwave background (e.g.~Hamilton
1997b; Knox, Bond \& Jaffe 1997; Tegmark \& Hamilton 1997).
In our case, a first step would consist in finding
estimators $\tA_k$ for which the covariance matrix $\langle \tA_k
\tA_{k'} \rangle - A_k A_k'$ is zero (SBC). 
Even if this is probably
feasable, since the calculation of the components of this matrix is similar
to what we did for the cosmic error, this constraint is not
sufficient, again because of the non-Gaussian nature of the underlying
statistics. Unfortunately, the calculation of 
higher order covariance matrixes such as $\langle \tA_k \tA_{k'} \tA_{k''}
\rangle - A_k A_{k'} A_{k''}$ looks quite tedious. 
\end{enumerate}

This paper constitutes the second step in a major investigation
on the theoretical errors on counts in cells. SC laid
the groundwork for all subsequent calculations, and here, their
formalism was extended and applied for magnitude limited redshift
surveys with realistic selection function. We proposed a new
set of estimators for the factorial moments, which includes
compensation for the effects of the selection function and
a minimum variance weighting. The integral equation for this
weight was solved, an excellent approximation found, and 
the corresponding errors were calculated. Optimal
sparse sampling strategies were considered as well, and it was shown that
in most cases the decrease in variance does not outweigh
the disadvantages of the corresponding information loss.
Therefore full sampling is advocated for most applications
requiring high order statistics.
Finally, the question of optimal survey geometry was addressed,
and we found that quasi random distribution of fields is a reasonable
choice, when a small fraction of the sky is covered by the survey.

\section*{Acknowledgments}
We thank A. Stebbins, 
D. Pogosyan, G. Mamon, J. Loveday, B. Guiderdoni, 
F.R. Bouchet, J.R. Bond, F. Bernardeau, and particularly
N. Kaiser and L. Knox for helpful discussions. I.S. is supported by
DOE and NASA through grant NAG-5-2788 at Fermilab. 
%
%

%
%
%
\appendix
%
%
%
\section[]{Calculation of the Cosmic Error}
%
%
This section generalizes the calculation of the cosmic error by SC to
the case of a non-uniform selection function, and 
for our estimator which includes local (spatial) statistical weight.

Following CBSII and SC, equation (\ref{eq:ecosm}) is separated into two
parts, according to whether the cells overlap or not
\begin{equation}
  E=E_{\rm overlap}+E_{\rm disjoint},
\end{equation}
with
\begin{equation}
  E_{\rm overlap} \equiv \frac{1}{\hat V} \int_{r_{12} 
  \leq 2\ell} d^3 r_1 d^3r_2 \ldots,
\end{equation}
\begin{equation}
  E_{\rm disjoint} \equiv \frac{1}{\hat V} \int_{r_{12} 
  \geq 2\ell} d^3 r_1 d^3r_2 \ldots,
\end{equation}
and $r_{12} \equiv |\r_1 - \r_2|$. Except when specified, we assume
that the cell size is small compared to the survey size. Our
calculations will thus be valid at leading order in $v/V$.  
%
\subsection{Contribution from Disjoint Cells}
%
The contribution to the error from disjoint cells corresponds to {\em
finite volume effects}. It requires the knowledge of the generating
function $P_{\r_1,\r_2}(x,y)$ of the bivariate counts in disjoint cells
at positions $\r_1$ and $\r_2$. The problem is that the average number
density of galaxies depends here on the distance $r$ from the
observer. According to the the reasoning of Szapudi \& Szalay (1993a)
in terms of Poisson processes [see, e.g., their eq.~(5.2)],  
\begin{eqnarray}
	\renewcommand{\baselinestretch}{1.0}
        P_{\r_1,\r_2}(x,y) &=& \displaystyle 
	\exp\big\{ 
        \sum_{N,M} {(x-1)^N(y-1)^M {\bar N}_{r_1}^N {\bar
        N}_{r_2}^M \over N! M! v^{N+M}} \nonumber \\
               &  &\displaystyle  
                   \int_{v_1}d^3u_1\ldots d^3u_N\int_{v_2}d^3u_{N+1}
                   \ldots d^3u_{N+M} \nonumber \\
	           & & \displaystyle
                   \xi_{N+M}(\u_1,\ldots ,\u_{N+M}) 
               \big\},
                   \label{biva}
\end{eqnarray}
where ${\bar N}_{r}$ is the average number of objects per cell
expected at position $\r$ [eq.~(\ref{eq:nbarn})]. There are two
approximations in SC which are used for the bivariate generating function.
We refer to SC for details about the underlying hypotheses in these
approximations, which are particular cases of the {\em hierarchical
model} (e.g.~Peebles 1980, BS), and for those it is trivial to take into
account the above supplementary dependence of the average densities
${\bar N}_{r_1}$ and ${\bar N}_{r_2}$ on distance from the
observer. The first approximation, hereafter SS, was derived by
Szapudi \& Szalay (1993a, 1993b). It assumes that the integral in
equation (\ref{biva}) can be well approximated as ${\bar N}_{r_1}^N
{\bar N}_{r_2}^M Q_{N+M} \Gamma_N(r_1) \Gamma_M(r_2) NM \xi$ up to
linear order in $\xi/\xiav$. The $r$ dependence (through ${\bar N}$)
of quantity $\Gamma_N$ [defined by eq.~(\ref{eq:gammaN})] is now
explicitly written, thus
\begin{eqnarray}
  P_{\r_1,\r_2}(x,y)  \simeq  P_{r_1}(x)P_{r_2}(y)[1+R_{\r_1,\r_2}(x,y)]
	\nonumber \\
      \mbox{}  + {\cal O}(\xi^2/\xiav^2),
\end{eqnarray}
\begin{eqnarray}
  R_{\r_1,\r_2}(x,y)  = \xi \sum_{M,\ N=1}^{\infty} (x-1)^N (y-1)^M
  Q_{N+M} \nonumber \\
   \Gamma_M(r_1) \Gamma_N(r_2) NM,
  \label{eq:SS}
\end{eqnarray}
where $\xi = \xi(r_{12})$.
The second approximation, hereafter BeS, was proposed by Bernardeau \&
Schaeffer (1992). It is
\begin{eqnarray}
  R_{\r_1,\r_2}(x,y) = \tau\left[ (1-x) {\bar N}(r_1) \xiav \right]
   \nonumber \\
   \tau\left[ (1-y) {\bar N}(r_2) \xiav \right] \xi/\xiav^2,
  \label{eq:BeS}
\end{eqnarray}
where
\begin{equation}
   \tau(s) = s \sqrt{2\sum_{N\ge 2} (-s)^{N-2} Q_N \frac{N^{N-2}
   (N-1)}{N!}}.
\end{equation}
It was noted by SC that the two approximations SS and BeS, although
quite different formally, give practically identical results for the
cosmic error on the factorial moments in realistic cases. Also, as
shown by SC, in the case of a sample without selection effects and 
with homogeneous weighting ($\phi=\omega=1$), the relative finite
volume error on $F_k$ does not depend on ${\bar N}$, neither on the average
density. This means, as shown later, that in the more general case
$\phi(r) \leq 1$, all $\phi$-dependent terms disappear in the final
expression for the finite volume error on $F_k$. 
%
\subsection{Contribution from Overlapping Cells}
%
The overlapping contribution is discussed under the assumption that the 
variations of $\phi(r)$ and $\omega(r)$ are small across the length
$2\ell$. This is a reasonable approximation when the cell size $\ell$
is small enough. With the additional assumption of local Poisson
behavior (SC)
\begin{eqnarray}
  E_{\rm overlap}(x,y) \simeq \frac{1}{\hat V} \int_{\hat V} d^3r 
  \omega^2(\r) [\phi(r)]^{-2k} \nonumber \\
  \left. E_{{\rm overlap}}^{\phi=\omega=1}(x,y) \right|_{{\bar
  N}={\bar N}_r},
  \label{eq:eover}
\end{eqnarray}
where the quantity $E_{\rm overlap}^{\phi=\omega=1}(x,y)$ is the
overlapping contribution to the generating function of the cosmic
error computed by SC when there are no selection effects and the
weighting function is unity. There is an $r$ dependence accounting
explicitly for the fact that the average number density $n \phi(r)$
depends on the distance from the observer: $E_{\rm
overlap}^{\phi=\omega=1}(x,y)$ is calculated at ${\bar N} = {\bar
N}_r$. The overlapping contribution of the error can be thus inferred
from the calculations of SC with a supplementary  weighted integral.  
%
\subsection{Cosmic Error on Factorial Moments}
%
Let us now concentrate on the error on the factorial moments, obtained
from equation (\ref{eq:facer}).

The finite volume contribution from disjoint cells is
\begin{equation}
   \left( \Delta^{\rm F} \tF_k \right)^2 = \frac{1}{{\hat V}^2} 
   \int_{r_{12} \geq 2\ell}
   d^3r_1 d^3 r_2 \omega(\r_1) \omega(\r_2)  \eta(\r_1,\r_2)
\end{equation}
with
\begin{eqnarray}
   \eta(\r_1,\r_2) & = & 
   [\phi(r_1)\phi(r_2)]^{-2k} 
   \big[ \frac{\partial}{\partial x} \big]^k \big[
   \frac{\partial}
   {\partial y} \big]^k \nonumber \\
   & & \left\{ P_{r_1}(x+1)P_{r_2}(y+1) \right. \nonumber \\
   & & \left.\left. R_{\r_1,\r_2}(x+1,y+1) \right\}\right|_{x=y=0}.
\end{eqnarray}
From equations (\ref{eq:genpn}) and (\ref{eq:SS}) or (\ref{eq:BeS}), 
function $\eta$ 
depends only on $r_{12}$, as follows
\begin{equation}
   \eta(\r_1,\r_2) \propto \xi(r_{12}).
\end{equation}
Thus the selection effects {\em cancel out} in the finite volume 
contribution to the cosmic error on the factorial moments. 
The results of SC (see their \S~4.3) can therefore be directly 
used here, just by replacing
their $\xiav(L)\equiv V^{-2} \int_V d^3r_1 d^3r_2 \xi(r_{12})$ 
by the weighted average
\begin{eqnarray}
   \xiav_{\rm eff}({\hat L}) & = & \frac{1}{{\hat V}^2} 
   \int_{r_1, r_2 \in \tV, r_{12} \geq
   2\ell} d^3 r_1 d^3 r_2 \omega(\r_1) \omega(\r_2)
   \xi(r_{12}).\nonumber \\
   & & \label{eq:xiaveff}
\end{eqnarray}
The quantity ${\hat L}$ stands for the effective catalogue size 
${\hat L} = {\hat V}^{1/3}$.
At leading order to $v/V$ the integral
(\ref{eq:xiaveff}) reduces to 
\begin{equation}
   \xiav_{\rm eff}({\hat L}) \simeq {\hat V}^{-2} 
   \int_{\hat V} d^3 r_1 d^3 r_2 \omega(\r_1) 
   \omega(\r_2) \xi(r_{12}).
\end{equation}
For $\omega=1$, naturally $\xiav_{\rm eff}({\hat
L})=\xiav({\hat L})$.
Thus the finite volume error with the notation of \S~3.2 is, 
\begin{eqnarray}
   \Delta^2_{\rm F}[\omega] & \equiv & 
   \left( \frac{\Delta^{\rm F} \tF_k}{F_k} \right)^2 \simeq
   \left( \frac{\Delta^{\rm F}_{\phi=\omega=1} \tF_k}{F_k}
   \right)^2 
   \frac{\xiav_{\rm eff}({\hat L})}{\xiav(L)} \nonumber \\
   & = & \Delta^2_{\rm
   F}\frac{\xiav_{\rm eff}({\hat L})}{\xiav(L)}.
\end{eqnarray}

The calculation of the contribution $(\Delta^{\rm overlap}\tF_k)^2$
from overlapping cells, the sum of the edge, and
the discreteness errors (see \S~3.2), is easier. Using the
fact that  ${\bar N}_r \propto \phi(r)$, $F_k \propto {\bar N}^k$ and
that  the relative edge effect error $\Delta^2_{\rm E}$ does not
depend on ${\bar N}$ in the case $\phi=\omega=1$, it is
straightforward to obtain equations (\ref{eq:deledge}) and
(\ref{eq:deldis}).
%
%
\section[]{Numerical Calculations}
%
%
\subsection{Cosmic Error and Minimum Variance Weight}
%
In this section the numerical solution of integral equation
(\ref{eq:optiwsph}) is considered. Let us define the step $\Delta$ as
\begin{equation}
  \Delta\equiv ({\hat R}_{\rm max}-{\hat R}_{\rm min})/N_{\rm bin},
\end{equation}
and the numbers 
\begin{equation}
  r_i={\hat R}_{\rm min} + (i-1/2)\Delta, \quad i=1,\ldots,N_{\rm
  bin}.
\end{equation}
The solution is obtained in the space of step functions $\omega$ verifying
\begin{equation}
  \omega(r)=\omega_i,\quad r\in [r_i-\Delta/2,r_i+\Delta/2[.
\end{equation}
Equation (\ref{eq:optiwsph}) in its discretized form can be written as
\begin{equation}
   \sum_{j=1}^{N_{\rm bin}} {\breve{\xi}}_{i,j} \omega_j + 
   \breve{\alpha}_i \omega_i + \breve{\lambda}_i=0, 
   \label{eq:mat}
\end{equation}
where
\begin{eqnarray}
   {\breve{\xi}}_{i,j} & 
   \equiv & \displaystyle\frac{\Delta^2_{\rm F}}{{\bar
   \xi}(L)
   {\hat V}}\int_{r_i-\Delta/2}^{r_i+\Delta/2}
   \int_{r_j-\Delta/2}^{r_j+\Delta/2} \nonumber \\
   & & \displaystyle u^2 {\hat \Omega}(u)
   v^2 {\hat \Omega}(v) {\tilde \xi}(u,v) du dv, 
   \label{eq:xi2cup}
\end{eqnarray}
\begin{equation}
   {\breve{\alpha}}_i=\int_{r_i-\Delta/2}^{r_i+\Delta/2} \left(
   \Delta^2_{\rm E}
   +\Delta^2_{\rm D}(u) \right) u^2 {\hat \Omega}(u) du,
\end{equation}
\begin{equation}
  \breve{\lambda}_i=\lambda \int_{r_i-\Delta/2}^{r_i+\Delta/2} u^2
  {\hat \Omega}(u) du.
\end{equation}
Note that the cosmic error is simply
\begin{equation}
   \Delta^2_{\rm cosmic}[\omega]=\sum_{i,j=1}^{N_{\rm bin}}
   \breve{\xi}_{i,j} \omega_i
   \omega_j + \sum_{i=1}^{N_{\rm bin}} \breve{\alpha}_i \omega_i^2.
   \label{eq:com}
\end{equation}
The most difficult part in determining the optimal sampling vector
$(\omega_i)_{i=1,\ldots,N_{\rm bin}}$ is the calculation of the
correlation matrix ${\breve{\xi}}_{i,j}$. A new binning is necessary
for computing the double integral (\ref{eq:xi2cup}). We proceed as
follows: 
\begin{eqnarray}
   {\breve{\xi}}_{i,j} & \simeq & \frac{\Delta^2_{\rm F}}{2{\bar
   \xi}(L){\hat V}} \sum_{k,l=1}^{N_{\rm subbin}} \left[ u_k^2 {\bar
   v}_l^2 {\hat \Omega}(u_k){\hat \Omega}({\bar v}_l) {\tilde
   \xi}(u_k,{\bar v}_l) \Delta u_k \Delta {\bar v}_l \right. \nonumber \\
   & & \mbox{} + \left. {\bar u}_k^2 v_l^2 {\hat \Omega}({\bar
   u}_k) {\hat \Omega}(v_l) {\tilde
   \xi}({\bar u}_k,v_l) \Delta {\bar u}_k \Delta v_l \right],
   \label{eq:xiii}
\end{eqnarray}
where
\begin{eqnarray}
   \Delta u_k & \!\!\!\!\!= & \!\!\!\!\! \Delta v_k=\left\{ \begin{array}{ll} 
		       \Delta(N_{\rm subbin}-1), & 
		       2 \leq k \leq N_{\rm subbin}-1 \\
		       \frac{1}{2}\Delta (N_{\rm subbin}-1), &
                       k=1\ {\rm or}\ N_{\rm subbin},
		      \end{array}
	      \right. \nonumber \\
   \Delta {\bar u}_k & \!\!\!\!\! = &\!\!\!\!\! 
   \Delta {\bar v}_k=\Delta/N_{\rm subbin},
\end{eqnarray}                     
\begin{eqnarray}
   u_k & = & r_i+(k-1)\Delta/(N_{\rm subbin}-1),\nonumber \\
   v_l & = & r_j+(l-1)\Delta/(N_{\rm subbin}-1), \nonumber \\
   {\bar u}_k & = & r_i+(k-1/2)\Delta{\bar u}_k, \nonumber \\
   {\bar v}_l & = & r_j+(l-1/2)\Delta{\bar v}_l.
\end{eqnarray}
The double summation in equation (\ref{eq:xiii}) 
avoids calculation of function ${\tilde \xi}(u,v)$ for $u=v$, where it
might diverge (for example, if $\xi(r) \propto r^{-\gamma}$ at small
scales, then function ${\tilde \xi}(u,u)$ diverges when $\gamma\geq
2$). It remains to compute function ${\tilde \xi}(u,v)$, $u \neq v$. In
equation (\ref{eq:defxit}), one can define $\theta$ as the angle
between the directions ${\hat r}_1$ and ${\hat r}_2$:
\begin{equation}
   \mu\equiv \cos(\theta)=\cos(\varphi_1-\varphi_2)\sin\theta_1
   \sin\theta_2 + \cos\theta_1 \cos\theta_2.
\end{equation}
Equation (\ref{eq:defxit}) can then be rewritten as
\begin{equation}
   {\tilde \xi}(u,v)=\int_{-1}^1 P(\mu|u,v) d\mu \xi\left(
   [u^2 + v^2 - 2 u v \mu ]^{1/2} \right),
   \label{eq:angulav}
\end{equation}
where $P(\mu|u,v)$ is the probability distribution function of $\mu$,
given $u$ and $v$. (For a spherical catalog 
$P(\mu|u,v)=1/2$). Typically, if $u$ and $v$ are large enough compared
to ${\hat R}_{\rm min}$, the function $P(\mu|u,v)$ does not depend sensitively
on the values of $u$ and $v$. It increases with $\mu$, from zero for
$-1 \leq \mu \leq \mu_{\rm min}$, where $\mu_{\rm min}$ corresponds to
the largest effective angular size of the catalog, to some maximum at
$\mu=1$, $P_{\rm max}$, depending on the angular size of the catalog.
For our SDSS-like catalog, $P_{\rm max}$ is of order 2. The
calculation of $P(\mu|u,v)$ is done numerically for a discrete
set of values of $(\mu_i,r_j,r_k)$, $1\leq i \leq N_{\rm \mu}$, $0
\leq j,k \leq N_{\rm bin}+1$ and $r_0 \equiv {\hat R}_{\rm min}$,
$r_{N_{\rm bin}+1} \equiv {\hat R}_{\rm max}$. For each value of
$(j,k)$, a Monte-Carlo simulation is done, i.e., directions ${\hat
r}_j$ and ${\hat r}_k$ are randomly chosen such that the corresponding
cells are included in the catalog. A reasonably accurate calculation,
with a few percent absolute errors requires $N_{\rm iter} \sim
100,000$ iterations for $N_{\mu}=100$. For a given
value of $(\mu,u,v)$, the estimate $P_{\rm interpol}(\mu,u,v)$ of
$P(\mu,u,v)$ is obtained from a bilinear interpolation between
$P(\mu_i,r_j,r_k)$, $P(\mu_i,r_{j+1},r_k)$, $P(\mu_i,r_j,r_{k+1})$,
$P(\mu_i,r_{j+1},r_{k+1})$ where $\mu_i$ is as close as possible to
$\mu$, and $r_j \leq u \leq r_{j+1}$, $r_k \leq v \leq r_{k+1}$. By
definition, we take $P(\mu_i,r_j,r_k)\equiv 0$ if $j=0$ or $k=0$.

We use the same Monte-Carlo simulation as above 
(except that we use only one direction ${\hat r}_j$) to compute the
effective solid angle on an array ${\hat \Omega}(r_j)$, and then
proceed with linear interpolations to compute estimates ${\hat
\Omega}_{\rm interpol}(u)$ of ${\hat \Omega}(u)$ at $u \neq r_j$. 

To compute the angular average (\ref{eq:angulav}) the following
variable is defined
\begin{equation}
   z(\mu)=\frac{1}{(\gamma_z-2)u v} \left( u^2 + v^2 - 2 u v \mu
   \right)^{1-\gamma_z/2},
   \label{eq:cople}
\end{equation}
where $\gamma_z \simeq \gamma$ is the expected power-law index of the
two-point correlation function at small scale, or some value close to
it. With the definitions $\mu(z)\equiv z^{-1}(z(\mu))$,
$z_{\rm m}=z(\mu=-1)$, $z_{\rm M}=z(\mu=1)$, 
\begin{equation}
   \Delta z_i=\left\{ \begin{array}{ll} 
		       (z_{\rm M}-z_{\rm m})/(N_z-1),\quad & 
		       2 \leq i \leq N_z-1, \\
		       \frac{1}{2}(z_{\rm M}-z_{\rm m})/(N_z-1), &
                       i=1\ {\rm or}\ N_z,
		      \end{array}
	      \right.
\end{equation}  
\begin{equation}
   z_i=z_m+(i-1)(z_{\rm M}-z_{\rm m})/(N_z-1),
\end{equation}
the correlation matrix can be written as
\begin{eqnarray}
  {\tilde \xi}(u,v) & \simeq & \sum_{i=1}^{N_z} \left[ u v (\gamma_z -2 )
  z_i \right]^{\gamma_z/(2-\gamma_z)} P_{\rm interpol}(\mu(z_i),u,v)
  \nonumber \\
  & & \xi\left( \left[ u v (\gamma_z
  -2 ) z_i \right]^{1/(2-\gamma_z)} \right) \Delta z_i.
  \label{eq:xiz}
\end{eqnarray}

The calculation of ${\breve{\alpha}}_i$ follows naturally from
\begin{equation}
   {\breve{\alpha}}_i=\sum_{k=1,N_{\rm subbin}} \!\!\!\! \left(\Delta^2_{\rm
   edge}+\Delta^2_{\rm D}(u_k)\right) u_k^2 {\hat \Omega}_{\rm
   interpol}(u_k) \Delta u_k,
\end{equation}
and one proceed similarly for $\breve{\lambda}_i$:
\begin{equation}
   \breve{\lambda}_i= \lambda \sum_{k=1,N_{\rm subbin}} u_k^2 {\hat
   \Omega}_{\rm interpol}(u_k) \Delta u_k.
   \label{eq:lamcal}
\end{equation}
The numerical estimate of the effective sample volume reads
\begin{eqnarray}
  {\hat V} & \simeq & \sum_{i=1}^{N_{\rm bin}} \frac{1}{r_{i+1}-r_{i}} 
  \left[ \left( {\hat \Omega}(r_{i+1})-{\hat \Omega(r_{i})} \right)
  \left( r_{i+1}^4 - r_i^4 \right)/4 \right. \nonumber \\
  & & + \left. \left({\hat \Omega(r_{i})} 
  r_{i+1} -{\hat \Omega}(r_{i+1}) r_i\right) \left( r_{i+1}^3 -r_i^3
  \right)/3 \right].
\end{eqnarray}
The same integration scheme is used to normalize the weight 
[eq.~(\ref{eq:omnorm})]. To compute the sample volume $V$ of our SDSS
like catalog, a more accurate estimator is used, with exact calculation
of $\Omega(r)$ and an integral on variable $r$ using the trapezoidal
method with $100,000$ points.

For the calculation of the terms $\Delta^2_{\rm F}$,
$\Delta^2_{\rm E}$ and $\Delta^2_{\rm D}$, we use the
results of SC corresponding to the SS approximation and to
$\gamma=1.8$ where $\gamma$ is the assumed logarithmic slope of
function $\xi(r)$ for computing the numerical coefficients in the
different terms contributing to the cosmic error [see equations (53)
to (68) in SC].  Note that, as discussed in SC, these numerical
coefficients are quite insensitive to $\gamma$ if it stays
reasonably close  to $1.8$. Therefore the same expression is applied for
the cosmic error as a function of $Q_N$, ${\bar N}$ and ${\bar \xi}$
even if $\gamma\neq1.8$ or if function ${\bar \xi}(\ell)$ is not
exactly a power-law, like in the CDM case. 
%
\subsection{Accuracy Tests}
%
The difficulty in obtaining accurate results for the optimal weight
relies mostly in the estimation of the correlation matrix
$\breve{\xi}_{i,j}$. Here, we assume that the finite volume error
dominates over the edge effect and the shot noise errors, although we
performed extensive accuracy tests including all sources of error. 

\begin{figure}
\centerline{\hbox{\psfig{figure=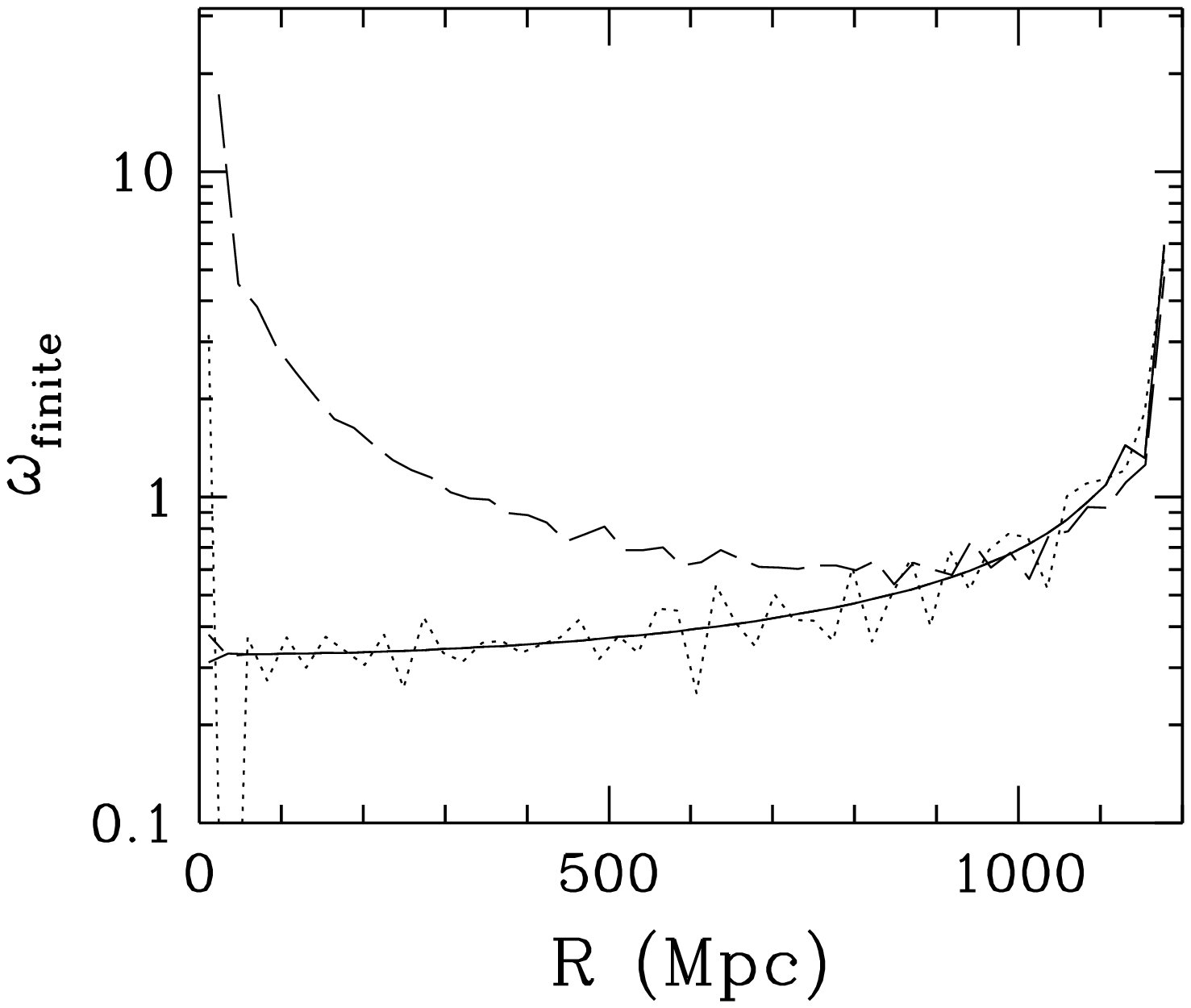,width=8cm}}}
\caption[]{The minimum variance weight $\omega$ is shown
as a function of distance
$R$ from the observer in the case the finite volume error is dominant
(the results displayed here correspond to $\ell=10$ Mpc, but would not
change significantly for other values of $\ell$). For the finite
volume error, $\omega$ does not depend on the order $k$.
The two point
function is assumed to be a power-law $\xi(r) \propto r^{-\gamma}$.
The calculations were performed assuming spherical geometry, 
except for the long dashes which show the result
for a SDSS-like geometry. The solid curve corresponds to the full
analytical calculation of the correlation matrix
$\breve{\xi}_{i,j}$. A dashed curve is almost perfectly superposing to
the solid one. In that case, the numerical scheme given by equations
(\ref{eq:xiii}), (\ref{eq:lamcal}) and (\ref{eq:xiz}) (with $P_{\rm
interpol}(\mu|u,v)\equiv 0.5$) has been used. The dotted line shows
the result as calculated via
a Monte-Carlo simulation plus bilinear interpolation to
compute $P_{\rm interpol}(\mu|u,v)$. The long dashes, corresponding to
a SDSS geometry, were obtained the same way as the dots.}
\label{fig:figure8}
\end{figure}

In what follows, we study the optimal weight $\omega(r)$, which
depends only on the shape of the two-point correlation function (and
thus not on higher order statistics). We consider a spherical sample,
where there is no need in principle to compute numerically
$P(\mu|u,v)$ as straightforwardly $P(\mu|u,v)=1/2$. Furthermore, the
two-point correlation function is assumed to be a power law of index
$-\gamma$ [eq.~(\ref{eq:xipow})], which allows the analytical
computation of the correlation matrix $\breve{\xi}_{i,j}$:
\begin{eqnarray}
  \breve{\xi}_{i,j}& =  \displaystyle \frac{8\pi^2 
  \Delta^2_{\rm F}}{{\bar
  \xi}(L){\hat V}} & \big[ F(r_i+\Delta/2,r_j+\Delta/2) \nonumber \\
 & &               \mbox{} +  F(r_i-\Delta/2,r_j-\Delta/2) \nonumber \\
 & &               \mbox{} -  F(r_i+\Delta/2,r_j-\Delta/2) \nonumber \\
 & &               \mbox{} -  F(r_i-\Delta/2,r_j+\Delta/2) \big]
   \label{eq:anaxi}
\end{eqnarray}
with
\begin{eqnarray}
     F(u,v) & \equiv & \displaystyle 
     \frac{r_0^{\gamma}}{(2-\gamma)(3-\gamma)(4-\gamma)(6-\gamma)} \nonumber \\
    & & \left[(6-\gamma) uv\left\{ |u-v|^{4-\gamma} + (u+v)^{4-\gamma} \right\}
     \right. \nonumber \\
    & & \mbox{} \left.  +\left\{ |u-v|^{6-\gamma} -
     (u+v)^{6-\gamma}\right\}
     \right],
\end{eqnarray}
and
\begin{equation}
  {\hat V}=\frac{4}{3} \pi {\hat R}_{\rm max}^3.
\end{equation}
Therefore
\begin{equation}
   \breve{\lambda}_i=4\pi r_i^2 \Delta 
   \left( 1 + \frac{\Delta^2}{12 r_i^2} \right).
   \label{eq:anala}
\end{equation}
In figure~\ref{fig:figure8}, the minimum variance weight is shown as a
function of distance from the observer, obtained from three different
calculations, each with $N_{\rm bin}=50$. The solid curve corresponds
to the analytical result [eqs.~(\ref{eq:anaxi}) and
(\ref{eq:anala})]. There is a dashed curve almost perfectly
superposing to the solid one, except from the extreme left part of
it. In that case, equations (\ref{eq:xiii}), (\ref{eq:lamcal}) and
(\ref{eq:xiz}) with $P(\mu|u,v)_{\rm interpol}=0.5$ were used to
compute $\breve{\xi}_{i,j}$ and $\breve{\lambda}_i$. We have taken
$N_{\rm subbin}=30$ and $N_z=50$ (as for all the figures of the main
text, when it is relevant). For both the solid and the dashed curves,
there is an irregularity at each end. The left irregularity,
corresponding to the weight given (nearly) at the center of the
catalog, is due to the limitations of the discrete approach. In the
continuous limit, the function $\omega(r)$ is expected to behave
smoothly in the neighborhood of $r=0$. The right irregularity is more
difficult to understand, and will be discussed later. The fluctuating
dotted curve is the same as the dashed one, but we computed the
probability distribution function $P(\mu|u,v)$ by Monte-Carlo
simulation as explained in \S~B.1 (with $N_{\rm iter}=100,000$
iterations for $N_{\mu}=100$). The left irregularity for this is quite
dramatic, as there is one value of the estimated weight which is
negative. Note, however, that the calculation of the finite volume
error is fairly robust with respect to the weight: the weights given
by the solid, the (invisible) short-dashed or the dotted curve give
exactly the same result. The long-dashed curve is the same as the
dotted one, expect that it corresponds to an SDSS like geometry
covering thus one quarter of the sky. The irregularities are less
pronounced in that case, due to the anisotropy of the distribution
$P(\mu|u,v)$ of values of $\mu$ (see also the solid curves of
Fig.~\ref{fig:figure1}).  

\begin{figure}
\centerline{\hbox{\psfig{figure=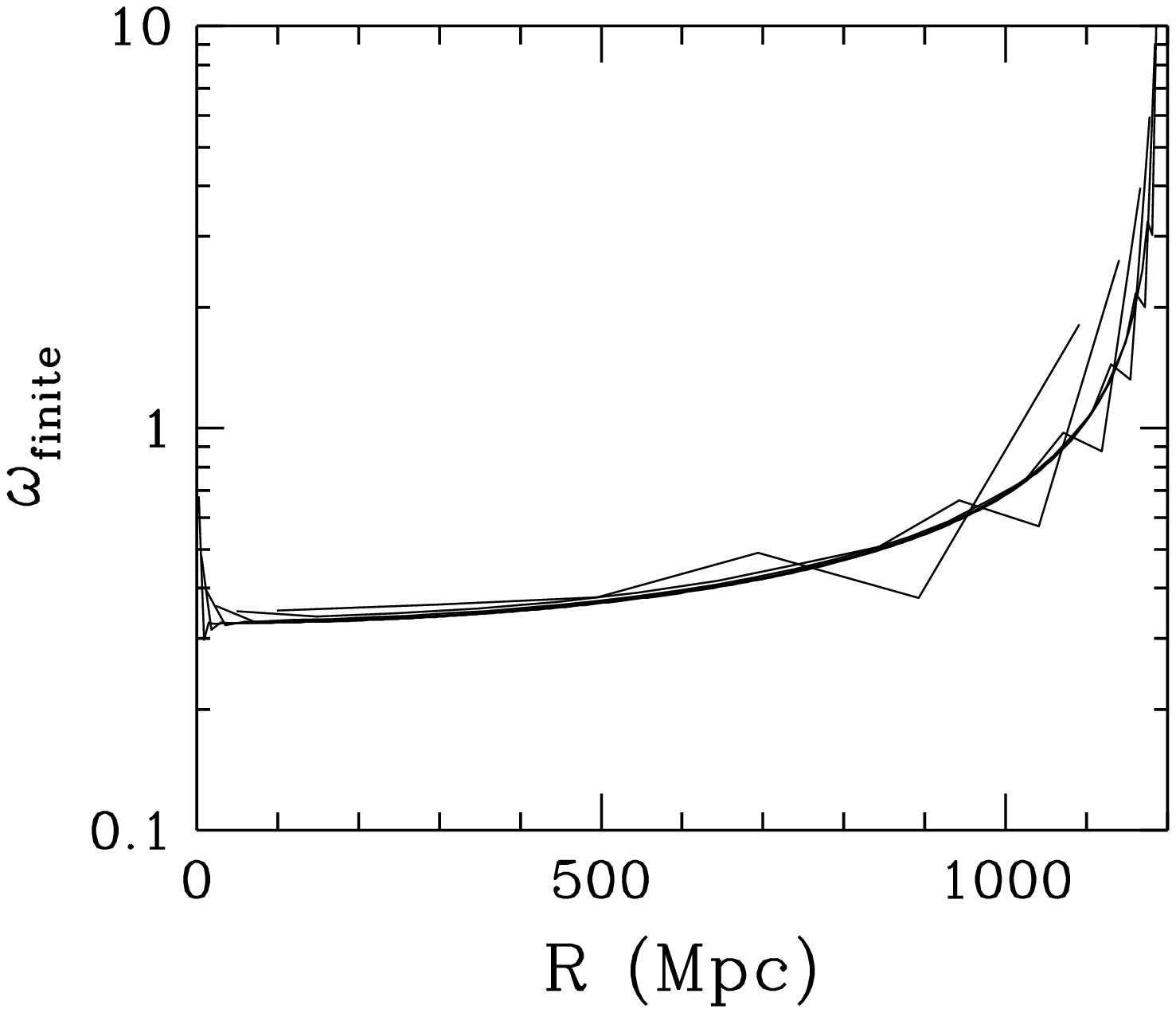,width=8cm}}}
\caption[]{Same as figure~\ref{fig:figure8} except that the
sensitivity of the results to the value of $N_{\rm bin}$
are tested for $N_{\rm bin}=6$, $12$, $25$, $50$, $100$ and $200$. The
numerical scheme given by equations (\ref{eq:xiii}), (\ref{eq:lamcal})
and (\ref{eq:xiz}) (with $P_{\rm interpol}(\mu|u,v)\equiv 0.5$) was
used.} 
\label{fig:figure9}
\end{figure}

Figure~\ref{fig:figure9} shows the effect of changing $N_{\rm
bin}$ when calculating the optimal  weight. There are six
curves, corresponding respectively to $N_{\rm bin}=6$, $12$, $25$,
$50$, $100$ and $200$. The method used for the calculation is the same
as the (almost invisible) dashed line in figure~\ref{fig:figure8}.
Except for the right and the left end-points of the curves, 
resolution is not critical for determining $\omega(r)$. The left
irregularity on each curve was shown to be unphysical and due to the
numerical limitations of our calculation. The right one is more of a
concern, since the estimator of $\log_{10}[\omega({\hat R}_{\rm max})]$ 
increases linearly with $N_{\rm bin}$, suggesting that the actual
optimal weight is singular at ${\hat R}_{\rm max}$. 
%
%
%
As a consequence, $N_{\rm bin}$ has to be large enough to resolve this
singularity sufficiently. The cosmic error can be sensitive to
the singularity, especially if the order $k$ is large and if the
finite volume error is significant. In realistic cases, however, the two-point
correlation does not behave like a power-law of index $-\gamma=-1.8$
up to arbitrarily large scales, thus the singularity is expected to be
less pronounced or to disappear. For example, in the CDM case, the effect is 
less significant, although there is still a slight instability at
${\hat R}_{\rm max}$ (see Fig.~\ref{fig:figure1}).  
%

%
\subsection{The optimal sampling rate}
%
The methods for generating figures~\ref{fig:figure6} and
\ref{fig:figure7} will be explained in this section. 
To simplify the calculations and since it was found
in \S~4.2 to be a good approximation, we take for the optimal
 weight $\omega$ the expression given by equation
(\ref{eq:goodapp}). To estimate the cosmic error from equation
(\ref{eq:com}), the matrix $\breve{\xi}_{i,j}$, and the vector
$\breve{\alpha}_i$ have to be computed for rather large values of
$R_{\rm max}$ in order to scan a large enough range of the sampling
rate $f$ [see eq.~(\ref{eq:rmax})]. Solving the implicit equation for
minimizing the cosmic error as a function of the sampling rate by
recalculating the matrix $\breve{\xi}_{i,j}$ for each value of $f$ 
would be too costly from the computational point of view. Instead, we
calculated $\breve{\xi}_{i,j}$ for $R_{\rm max}\equiv 5000$ Mpc and
with $N_{\rm bin}=200$. A given value of $f$ corresponds to some
${\hat R}_{\rm max}(f)$ and thus roughly to some value of $r_i$. This
is taken to be the as close as possible to ${\hat R}_{\rm max}(f)$. To
compute the cosmic error corresponding to some value of $f$, quadratic
interpolation is performed between the values obtained for ${\hat
R}_{\rm max}(f)=r_{i-1}$, ${\hat R}_{\rm max}(f)=r_{i}$ and ${\hat
R}_{\rm max}(f)=r_{i+1}$. Even this smoothing procedure cannot
guarantee smoothness for $f$ as a function of scale, although the
calculation is valid within a few percents. The implicit equation for
finding the minimum of function $\Delta^2_{\rm cosmic}(f)$ is easy to
solve by bisection as this function is convex (see
figure~\ref{fig:figure7}).
%
\section[]{Asymptotic Regimes}
%
Using the results of SC (valid at leading order in $v/V$), this
section presents an analytic estimate of the optimal sampling rate $f$
for a three-dimensional galaxy catalog (\S~5.1 and 5.2). This is
possible in the highly and weakly nonlinear regimes, where ${\bar
\xi}\gg 1$ and ${\bar \xi} \ll 1$, respectively. According to the
results of \S~5.1, uniform selection $\phi=1$ can be assumed. 

From the calculations of SC, the cosmic error on
the factorial moments of order $k\leq 3$ in the highly nonlinear
regime is (using the SS approximation, see \S~A.1)
\begin{equation}
 \left( \frac{\Delta_{\rm cosmic}^2 \tF_1}{F_1} \right)^2 \simeq {\bar
\xi}(L) +{\bar \xi}\frac{v}{V} \left[ \frac{1}{N_{\rm c}} +  5.51 \right],
 \label{eq:delf1}
\end{equation}
\begin{eqnarray}
 \left( \frac{\Delta_{\rm cosmic}^2 \tF_2}{F_2} \right)^2 
 \!\!\!\!\!\! & \simeq & \!\!\!\! 4 Q_4
 {\bar \xi}(L) + {\bar \xi}\frac{v}{V} \left[ \frac{0.5}{N_{\rm c}^2}
 \right. \nonumber \\
 & & \mbox{} \left. +\frac{6.6 Q_3}{N_{\rm c}} + 42.2 Q_4 \right],
\end{eqnarray}
\begin{eqnarray}
  \left( \frac{\Delta_{\rm cosmic}^2 \tF_3}{F_3} \right)^2 \!\!\!\!\!\!
  & \simeq & \!\!\!\!
  \frac{9Q_6}{Q_3^2} {\bar \xi}(L)+{\bar \xi}\frac{v}{V} \left[
  \frac{0.19}{Q_3} \frac{1}{N_{\rm c}^3}+\frac{4.93 Q_4}{Q_3^2}
  \frac{1}{N_{\rm c}^2}  \right. \nonumber \\ 
  & & \mbox{} \left.  + \frac{38.8}{Q_3^2} \frac{1}{N_{\rm c}} +
  \frac{185 Q_6}{Q_3^2} \right],
\end{eqnarray}
where
\begin{equation} 
  N_{\rm c} \equiv {\bar N}\ {\bar \xi}
\end{equation}
is the typical number of galaxies in a cell located in an overdense
region (e.g.~BS). At small enough scales and if $\gamma < 3$, it is 
possible that $N_{\rm c} \ll 1$ irrespectively of $f$. In that case,
count-in-cells statistics is dominated by discreteness effects
(BS). The edge effect error, corresponding to the term proportional to
${\bar \xi} v/V$ and independent of $N_{\rm c}$, is negligible
compared to the shot noise error. Equations (\ref{eq:foptima}) or
(\ref{eq:foptimabis}) follow straightforwardly. 

In the weakly nonlinear regime, where ${\bar \xi} \ll 1$, the Gaussian limit
at leading order in ${\bar \xi}$ is a good approximation. 
The expression (\ref{eq:delf1}) 
for the error on $F_1$ is still valid. At higher order, however, 
\begin{eqnarray}
\left( \frac{\Delta_{\rm cosmic}^2 \tF_2}{F_2} \right)^2 & \simeq &  4
{\bar \xi}(L) + 17.05 {\bar \xi} \frac{v}{V} \nonumber \\ 
& & \mbox{} +(0.65 + 4 {\bar N})
\frac{1}{{\bar N}^2} \frac{v}{V},
  \label{eq:delf2}
\end{eqnarray}
\begin{eqnarray}
  \left( \frac{\Delta_{\rm cosmic}^2 \tF_3}{F_3} \right)^2 & \simeq &
  9 {\bar \xi}(L) + 34.6 {\bar \xi} \frac{v}{V} \nonumber \\
  & & \mbox{} + \left(0.88 + 5.83 {\bar
N} + 9 {\bar N}^2 \right) \frac{1}{{\bar N}^3} \frac{v}{V}.
 \label{eq:delf3}
\end{eqnarray}
If the scale is large enough, edge effects are expected to dominate
over the finite volume errors, but this property depends on the
details of the underlying statistics and on the geometry of the
catalog. More rigorous estimates of the cosmic error, taking fully
into account the geometry of the catalog would be needed to prove it,
a level of accuracy outside of the scope of this paper.
Exact calculations of the cosmic error when the size of the cell
becomes comparable to the size of the catalog are indeed quite tedious
(see for example Appendix B of SC). Although our approach is valid
only to leading order in $v/V$, setting $v/V=1$ in equations
(\ref{eq:delf1}), (\ref{eq:delf2}), and (\ref{eq:delf3}) yields at
least a lower limit for the relative contribution of edge effects
compared to finite volume effects. The result is that the edge error
is a few times larger than the finite volume error (for a spherical
catalog). Assuming that edge effects dominate, equations
(\ref{eq:foptima2}) and (\ref{eq:foptima2bis}) naturally follow. 

Note that if finite volume effects dominate edge effects in the regime
${\bar \xi} \ll 1$, and ${\bar N} \gg 1$, the optimal sampling rate,
with the notations of \S~5, becomes 
\begin{equation}
   f=\left[ \frac{1-3\zeta}{\gamma_L \zeta N_{\rm ref}} \left(
   \frac{L_{\rm ref}}{L_0} \right)^{\gamma_L}
   \right]^{\frac{1}{1+(\gamma_L-3)\zeta}},
\end{equation}
where $\zeta=1/5$ or $1/9$ according to whether the redshifts are
collected individually (\S~5.1) or collectively (\S~5.2). For the same
choice of the parameters as in \S~5, we find $f\sim 1/20$ for
individual collection of redshifts and $f\sim 1/10$ for simultaneous
collection independently of scale and of the order $k$. As a result,
the general conclusions of our analysis in \S~5.1 and 5.2
are not changed, even if finite volume effects dominate over edge
effects.

\bsp

\label{lastpage}

%
\end{document}